%% file: main.tex
\documentclass[twocolumn,10pt]{IEEEtran}

\usepackage[T1]{fontenc}
\usepackage[latin9]{inputenc}
\usepackage{amsmath}
\usepackage{amssymb}
\usepackage{amsthm}
\usepackage{graphicx}
\usepackage{comment}
\usepackage{booktabs}
\usepackage[outdir=./]{epstopdf}
\usepackage[ruled,linesnumbered]{algorithm2e}
\usepackage[acronym]{glossaries}
\glsdisablehyper
\loadglsentries{glossary}
\newtheorem{proposition}{Proposition}
\usepackage{xcolor}

\let\oldnl\nl
\newcommand{\nonl}{\renewcommand{\nl}{\let\nl\oldnl}}

\DeclareMathOperator*{\argmin}{arg\,min}

\begin{document}

\title{Discrete-Value Group and Fully Connected Architectures for Beyond Diagonal Reconfigurable Intelligent Surfaces}

\author{Matteo Nerini,~\IEEEmembership{Graduate Student Member,~IEEE},
        Shanpu Shen,~\IEEEmembership{Senior Member,~IEEE},\\
        Bruno Clerckx,~\IEEEmembership{Fellow,~IEEE}

\thanks{Copyright (c) 2015 IEEE.
Personal use of this material is permitted.
However, permission to use this material for any other purposes must be obtained from the IEEE by sending a request to pubs-permissions@ieee.org.
This work was supported by Hong Kong Research Grants Council through the Collaborative Research Fund under Grant C6012-20G.
\textit{(Corresponding author: Shanpu Shen.)}}
\thanks{M. Nerini is with the Department of Electrical and Electronic Engineering, Imperial College London, London SW7 2AZ, U.K. (e-mail: m.nerini20@imperial.ac.uk).}
\thanks{S. Shen is with the Department of Electronic and Computer Engineering, The Hong Kong University of Science and Technology, Clear Water Bay, Kowloon, Hong Kong (e-mail: sshenaa@connect.ust.hk).}
\thanks{B. Clerckx is with the Department of Electrical and Electronic Engineering, Imperial College London, London SW7 2AZ, U.K., and with Silicon Austria Labs (SAL), Graz A-8010, Austria (e-mail: b.clerckx@imperial.ac.uk).}}


\maketitle

\begin{abstract}
Reconfigurable intelligent surfaces (RISs) allow controlling the propagation environment in wireless networks through reconfigurable elements.
Recently, beyond diagonal RISs (BD-RISs) have been proposed as novel RIS architectures whose scattering matrix is not limited to being diagonal.
However, BD-RISs have been studied assuming continuous-value scattering matrices, which are hard to implement in practice.
In this paper, we address this problem by proposing two solutions to realize discrete-value group and fully connected RISs.
First, we propose scalar-discrete RISs, in which each entry of the RIS impedance matrix is independently discretized.
Second, we propose vector-discrete RISs, where the entries in each group of the RIS impedance matrix are jointly discretized.
In both solutions, the codebook is designed offline such as to minimize the distortion caused in the RIS impedance matrix by the discretization operation.
Numerical results show that vector-discrete RISs achieve higher performance than scalar-discrete RISs at the cost of increased optimization complexity.
Furthermore, fewer resolution bits per impedance are necessary to achieve the performance upper bound as the group size of the group connected architecture increases.
In particular, only a single resolution bit is sufficient in fully connected RISs to approximately achieve the performance upper bound.
\end{abstract}

\glsresetall


\section{Introduction}

Reconfigurable intelligent surfaces (RISs), or intelligent reflecting surfaces, are an emerging technology that will enhance the performance of future wireless communications \cite{bas19}, \cite{wu19a}, \cite{liu21}.
This technology relies on large planar surfaces comprising multiple reflecting elements, each of them able to induce a certain amplitude and phase change to the incident electromagnetic wave.
Thus, an RIS can steer the reflected signal toward the intended direction by smartly coordinating the reflection coefficients of its elements.
RIS-aided communication systems benefit from several advantages.
RISs with passive elements are characterized by ultra-low power consumption and do not cause any active additive thermal noise or self-interference phenomena.
In addition, RIS is a low-profile and cost-effective solution, since it does not include \gls{rf} chains.

To avoid difficult optimization problems, many studies on RISs do not pose limitations on the allowed reflection coefficient values.
However, in practical implementation, they are selected from a finite number of discrete values.
Reflection coefficients tunable with finer resolution require a more complex hardware design, which can be prohibitive when the number of RIS elements is high \cite{wu21}.
Considering a single-user RIS-aided \gls{siso} system, the effects of discrete phase shifts have been investigated on the diversity order \cite{bad19}, \cite{xu21}, on the achievable rate \cite{zha20b}, and on the ergodic capacity \cite{li20}.
Furthermore, in \cite{liu20,an21,wei21,wan21},
channel estimation strategies have been proposed for RIS-aided systems.
In \cite{an22}, a codebook-based framework is studied to strike flexible trade-offs between communication performance and signaling overhead.
%
Discrete reflection coefficients have been optimized to enhance the performance of \gls{mimo} communications by solving rate maximization problems in single-user systems \cite{xu19}, \cite{you20}, \cite{abe20}, \cite{qi20} and sum-rate maximization problems in multi-user systems \cite{guo19}, \cite{di20}, \cite{mu20}, \cite{jun21}, \cite{zha21}, \cite{zha21b}.
In \cite{wu19b}, \cite{fu21}, the transmit power is minimized by jointly optimizing the continuous transmit precoding and the discrete phase shifts in the RIS.
For RIS-aided downlink communications, works have been conducted to maximize energy efficiency \cite{hua18}, \cite{hua19}, and the spectral efficiency in the presence of statistical \gls{csi} \cite{han20}.
In addition, RISs with discrete reflection coefficients have been investigated for Terahertz (THz) communications \cite{ma20a}, \cite{ma20b}, \cite{che19}, \cite{lu20}, wideband communications \cite{cai20}, \gls{swipt} \cite{wu20}, \cite{zha20}, \cite{gon21} and index modulation \cite{liu22}.
Finally, prototypes of discrete phase shift RISs have been designed in \cite{dai20}, \cite{dun20}.

\begin{figure*}[t]
    \centering
    \includegraphics[width=0.9\textwidth]{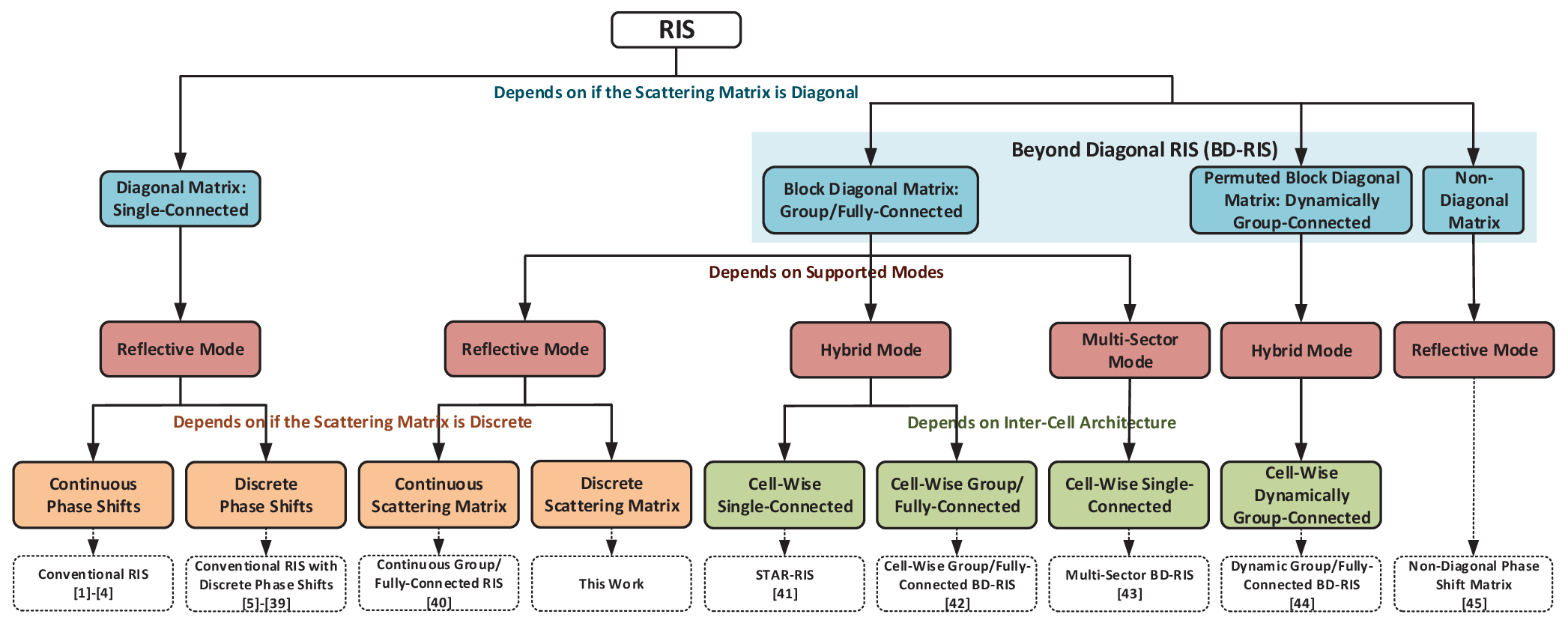}
    \caption{RIS classification tree.}
    \label{fig:ris-tree}
\end{figure*}

In the aforementioned literature \cite{bad19}-\cite{dun20}, it is always assumed that each RIS element is independently controlled by a tunable impedance connected to ground.
As a result, conventional RISs are characterized by a diagonal scattering matrix, also known as phase shift matrix.
This traditional RIS architecture is denoted as single connected \cite{she20}.
Recently, beyond diagonal RISs (BD-RISs) have been proposed as a novel branch of RISs in which the scattering matrix is not limited to be diagonal \cite{she20}.
In \cite{she20}, the authors generalized the single connected architecture by connecting all or a subset of RIS elements through a reconfigurable impedance network, resulting in the fully and group-connected architecture, respectively.
In \cite{xu21b}, the concept of \gls{star-ris} has been introduced.
This RIS architecture is able to reflect and transmit the impinging signal, differently from conventional RISs working only in reflective mode.
In \cite{li22-1}, a general RIS model has been proposed to unify different modes (reflective/transmissive/hybrid) and different architectures (single/group/fully connected).
The authors also propose different inter-cell architectures, namely cell-wise group/fully connected RIS architectures, where part of/all the cells are connected to each other \cite{li22-1}.
In \cite{li22-2}, BD-RISs supporting multi-sector mode have been proposed to achieve full-space coverage.
In multi-sector BD-RISs, the antennas are divided into multiple sectors, with each sector covering a narrow region of space.
Multi-sector BD-RISs have important gains over \gls{star-ris} due to the highly directional beam of each sector.
In \cite{li22-3}, dynamically group connected BD-RIS are optimized based on a dynamic grouping strategy.
In addition, in \cite{li22}, an RIS architecture with non-diagonal phase shift matrix is proposed, able to achieve a higher rate than conventional single connected RISs.

The group and fully connected RISs proposed in \cite{she20} have been studied assuming continuous-value scattering matrices, and it is not clear how to design these architectures with discrete-value scattering matrices.
Discrete-value single connected RISs have been typically realized by considering uniform discretization of the phase shifts in the interval $[0,2\pi)$.
This is made possible by the diagonal structure of the scattering matrix of these RISs, which is completely described by its phase shifts.
Thus, this discretization design cannot be applied to group and fully connected RISs, whose scattering matrices are not diagonal.
This gap is addressed in this paper, where we propose a design strategy for discrete-value group and fully connected RISs.
In Fig.~\ref{fig:ris-tree}, we classify the studies on RIS to better contextualize our work.
The contributions of this paper are summarized as follows.

\textit{First}, we propose a strategy to design discrete-value group and fully connected RISs that assigns a finite number of bits to each entry of the RIS impedance matrix.
The resulting RISs are denoted as scalar-discrete RISs, since they are obtained with a scalar quantization approach.
Firstly, a suitable codebook is designed in the offline learning stage.
Secondly, the discrete elements of the impedance matrix are optimized based on the resulting codebook in the online deployment stage.

\textit{Second}, we prove that our codebook always achieves a received signal power growth of $\mathcal{O}(N_I^2)$ in group connected architectures, where $N_I$ is the number of RIS elements.
Thus, the proposed discretization strategy does not degrade the growth of the received signal power as a function of $N_I$.

\textit{Third}, we develop a different discretization strategy for group and fully connected RISs, exploiting the block diagonal structure of their impedance matrices.
According to this strategy, we jointly discretize and optimize the entries in each block of the RIS impedance matrix through vector quantization.
RISs designed with this strategy, denoted as vector-discrete RISs, achieve higher performance than scalar-discrete RISs at the cost of increased optimization computational complexity.
Vector discretization has never been considered for conventional RISs since their impedance matrix is diagonal.

\textit{Fourth}, we assess the performance in terms of received signal power of single, group, and fully connected RISs employing different numbers of resolution bits.
We verify that the more connections are present in the reconfigurable impedance network, the less resolution is needed to achieve the optimal performance of continuous-value RISs.
Because of this property, in the fully connected architecture, a single resolution bit allocated to each reconfigurable impedance is sufficient to achieve the performance of ideal RISs.

\textit{Organization}: In Section~\ref{sec:system-model}, we introduce the system model and the problem formulation.
In Sections~\ref{sec:scalar-discrete} and \ref{sec:vector-discrete}, we present our novel discrete group and fully connected RIS design based on scalar and vector quantization, respectively.
In Section~\ref{sec:results}, we assess the performance of \gls{mimo} systems aided by discrete RISs in terms of received signal power.
Finally, Section~\ref{sec:conclusion} contains the concluding remarks.

\textit{Notation}: Vectors and matrices are denoted with bold lower and bold upper letters, respectively.
Scalars are represented with letters not in bold font.
$\left|a\right|$, and $\arg\left(a\right)$ refer to the modulus and phase of a complex scalar $a$, respectively.
$\left[\mathbf{a}\right]_{i}$ and $\left\|\mathbf{a}\right\|$ refer to the $i$th element and $l_{2}$-norm of vector $\mathbf{a}$, respectively.
$\mathbf{A}^{T}$, $\mathbf{A}^{H}$, $\left[\mathbf{A}\right]_{i,j}$, and $\left\|\mathbf{A}\right\|$ refer to the transpose, conjugate transpose, $\left(i,j\right)$th element, and spectral norm of matrix $\mathbf{A}$, respectively.
$\mathbb{R}$ and $\mathbb{C}$ denote the real and complex number sets, respectively.
$j=\sqrt{-1}$ denotes the imaginary unit.
$\mathbf{0}$ and $\mathbf{I}$ denote an all-zero matrix and an identity matrix, respectively, with appropriate dimensions.
$\mathcal{CN}\left(\mathbf{0},\mathbf{I}\right)$ denotes the distribution of a circularly symmetric complex Gaussian random vector with mean vector $\mathbf{0}$ and covariance matrix $\mathbf{I}$ and $\sim$ stands for \textquotedblleft distributed as\textquotedblright.
diag$\left(a_{1},\ldots,a_{N}\right)$ refers to a diagonal matrix with diagonal elements being $a_{1},\ldots,a_{N}$.
diag$\left(\mathbf{A}_{1},\ldots,\mathbf{A}_{N}\right)$ refers to a block diagonal matrix with blocks being $\mathbf{A}_{1},\ldots,\mathbf{A}_{N}$.
ones$\left(N\right)$ refers to an $N\times N$ matrix whose elements are all ones.

\section{System Model}
\label{sec:system-model}

\begin{figure}[t]
    \centering
    \includegraphics[width=0.48\textwidth]{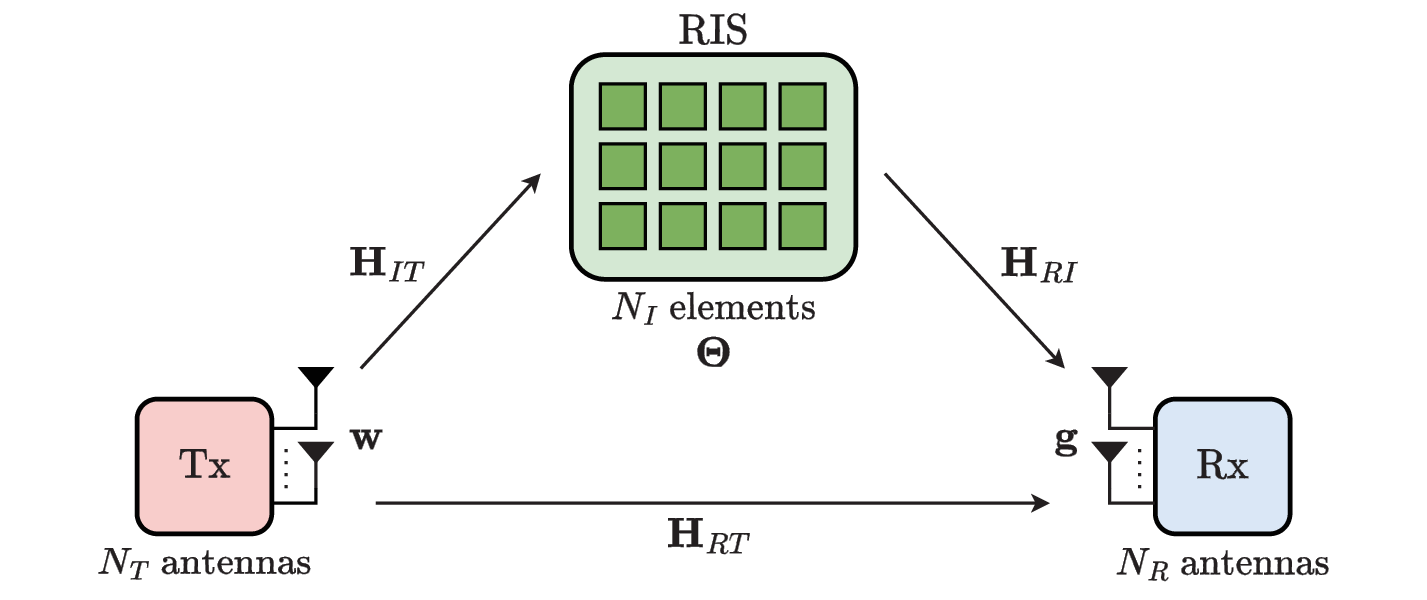}
    \caption{RIS-aided MIMO communication system model.}
    \label{fig:ris-system}
\end{figure}

Let us consider an RIS-aided \gls{mimo} system, as represented in Fig.~\ref{fig:ris-system}.
We denote as $N_{T}$ the number of antennas at the transmitter, $N_{R}$ the number of antennas at the receiver, and $N_{I}$ the number of antennas at the RIS.
The $N_{I}$ antennas of the RIS are connected to a $N_{I}$-port reconfigurable impedance network, with scattering matrix $\boldsymbol{\Theta}\in\mathbb{C}^{N_{I}\times N_{I}}$.
To characterize the channel matrix seen by the receiver $\mathbf{H}\in\mathbb{C}^{N_{R}\times N_{T}}$ as a function of $\boldsymbol{\Theta}$, we assume all antennas perfectly matched and we assume no mutual coupling between the antennas, as widely adopted in the literature \cite{she20}\footnote{In practice, these two assumptions can be achieved by individually matching each RIS antenna to the reference impedance $Z_{0}=50\:\Omega$ and by keeping the spacing between the RIS elements larger than half-wavelength.}.
With these two assumptions, the channel matrix $\mathbf{H}$ can be written as
\begin{equation}
\mathbf{H}=\mathbf{H}_{RT}+\mathbf{H}_{RI}\boldsymbol{\Theta}\mathbf{H}_{IT},\label{eq:H}
\end{equation}
where $\mathbf{H}_{RT}\in\mathbb{C}^{N_{R}\times N_{T}}$, $\mathbf{H}_{RI}\in\mathbb{C}^{N_{R}\times N_{I}}$, and $\mathbf{H}_{IT}\in\mathbb{C}^{N_{I}\times N_{T}}$ are the channels from transmitter to receiver, RIS to receiver, and transmitter to RIS, respectively \cite{she20}.
Perfect knowledge of the channels $\mathbf{H}_{RT}$, $\mathbf{H}_{RI}$, and $\mathbf{H}_{IT}$ is assumed, that can be acquired, for instance, through the semi-passive channel estimation protocol in \cite{wu21}.

We consider a single-stream \gls{mimo} transmission scheme, to exploit the diversity gain offered by the multiple antennas at the transmitter and receiver, and by the reflecting elements at the RIS.
We denote the transmit signal as $\mathbf{x}=\mathbf{w}s\in\mathbb{C}^{N_{T}\times1}$, where $\mathbf{w}\in\mathbb{C}^{N_{T}\times1}$ is the precoding vector subject to the constraint $\left\|\mathbf{w}\right\|=1$, and $s\in\mathbb{C}$ is the transmit symbol with average power $P_{T}=\mathrm{E}[\left|s\right|^{2}]$.
Denoting the receive signal as $\mathbf{y}\in\mathbb{C}^{N_{R}\times1}$, we have $\mathbf{y}=\mathbf{H}\mathbf{x}+\mathbf{n}$, where $\mathbf{H}$ is the \gls{mimo} channel matrix given by \eqref{eq:H}, and $\mathbf{n}\sim\mathcal{CN}\left(\mathbf{0},\sigma_{n}^{2}\mathbf{I}\right)$ is the \gls{awgn} at the receiver with variance $\sigma_{n}^{2}$.
Thus, the signal used for detection is given by $z=\mathbf{g}\mathbf{y}\in\mathbb{C}$, where $\mathbf{g}\in\mathbb{C}^{1\times N_{R}}$ is the combining vector subject to the constraint $\left\|\mathbf{g}\right\|=1$.
Eventually, we can express the signal $z$ as
\begin{equation}
z=\mathbf{g}\mathbf{H}\mathbf{w}s+\tilde{n},\label{eq:z}
\end{equation}
where $\tilde{n}=\mathbf{g}\mathbf{n}$ is the \gls{awgn} with variance $\sigma_{n}^{2}$.

The $N_{I}$-port reconfigurable impedance network is constructed with passive elements which can be adapted to the channel to properly reflect the incident signal.
To maximize the power reflected by the RIS, we consider the impedance matrix of the $N_{I}$-port reconfigurable impedance network $\mathbf{Z}_{I}\in\mathbb{C}^{N_{I}\times N_{I}}$ purely reactive.
Thus, we can write $\mathbf{Z}_{I}=j\mathbf{X}_{I}$, where $\mathbf{X}_{I}\in\mathbb{R}^{N_{I}\times N_{I}}$ denotes the reactance matrix of the $N_{I}$-port reconfigurable impedance network.
Hence, according to network theory \cite{poz11}, $\boldsymbol{\Theta}$ is given by
\begin{equation}
\boldsymbol{\Theta}=\left(j\mathbf{X}_{I}+Z_{0}\mathbf{I}\right)^{-1}\left(j\mathbf{X}_{I}-Z_{0}\mathbf{I}\right).\label{eq:T(X)}
\end{equation}
Furthermore, the reconfigurable impedance network is also reciprocal so that we have $\mathbf{X}_{I}=\mathbf{X}_{I}^{T}$ and $\boldsymbol{\Theta}=\boldsymbol{\Theta}^{T}$.
Depending on the topology of the reconfigurable impedance network, three different RIS architectures have been identified in \cite{she20}, which are described in the following.

\subsection{Single Connected RIS Architecture}

The single connected RIS architecture is the conventional architecture adopted in the literature \cite{bas19}, \cite{wu19a}, \cite{liu21}.
Here, each port of the reconfigurable impedance network is connected to ground with a reconfigurable impedance and is not connected to the other ports.
The reactance matrix $\mathbf{X}_{I}$ is a diagonal matrix given by $\mathbf{X}_{I}=\mathrm{diag}\left(X_{1},X_{2},\ldots,X_{N_{I}}\right)$, where $X_{n_{I}}$ is the reactance connecting the $n_{I}$th port to ground, for $n_{I}=1,\ldots,N_{I}$.
According to \eqref{eq:T(X)}, the scattering matrix $\boldsymbol{\Theta}$ is also a diagonal matrix written as
\begin{equation}
\boldsymbol{\Theta}=\mathrm{diag}\left(e^{j\theta_{1}},e^{j\theta_{2}},\ldots,e^{j\theta_{N_{I}}}\right),\label{eq:diag(T)}
\end{equation}
where $e^{j\theta_{n_{I}}}=\frac{jX_{n_{I}}-Z_{0}}{jX_{n_{I}}+Z_{0}}$ is the reflection coefficient of the reactance $X_{n_{I}}$, for $n_{I}=1,\ldots,N_{I}$.

\subsection{Fully Connected RIS Architecture}

The fully connected RIS architecture is obtained by connecting every port of the reconfigurable impedance network to all other ports.
Therefore, the reactance matrix $\mathbf{X}_{I}$ can be an arbitrary symmetric matrix.
According to \eqref{eq:T(X)}, $\boldsymbol{\Theta}$ is a complex symmetric unitary matrix
\begin{equation}
\boldsymbol{\Theta}=\boldsymbol{\Theta}^{T},\:\boldsymbol{\Theta}^{H}\boldsymbol{\Theta}=\boldsymbol{\mathrm{I}}.\label{eq:T fully}
\end{equation}

\subsection{Group Connected RIS Architecture}

The group connected RIS architecture has been proposed as a trade-off between the single connected and the fully connected to achieve a good balance between performance and complexity.
In the group connected architecture, the $N_{I}$ elements are divided into $G$ groups, each having $N_{G}=\frac{N_{I}}{G}$ elements.
Each element of the $N_{I}$-port is connected to all other elements in its group, while there is no connection inter-group. 
Thus, $\mathbf{X}_{I}$ is a block diagonal matrix given by
\begin{equation}
\mathbf{X}_{I}=\mathrm{diag}\left(\mathbf{X}_{I,1},\mathbf{X}_{I,2},\ldots,\mathbf{X}_{I,G}\right),\:\mathbf{X}_{I,g}=\mathbf{X}_{I,g}^{T},\:\forall g,\label{eq:X group}
\end{equation}
where $\mathbf{X}_{I,g}\in\mathbb{R}^{N_{G}\times N_{G}}$ is the reactance matrix of the $N_{G}$-port fully connected reconfigurable impedance network for the $g$th group.
According to \eqref{eq:T(X)}, the following constraints can be found for the scattering matrix in the group connected architecture
\begin{equation}
\boldsymbol{\Theta}=\mathrm{diag}\left(\boldsymbol{\Theta}_{1},\boldsymbol{\Theta}_{2},\ldots,\boldsymbol{\Theta}_{G}\right),\:\boldsymbol{\Theta}_{g}=\boldsymbol{\Theta}_{g}^{T},\:\boldsymbol{\Theta}_{g}^{H}\boldsymbol{\Theta}_{g}=\boldsymbol{\mathrm{I}},\:\forall g,\label{eq:T group}
\end{equation}
which show that $\boldsymbol{\Theta}$ is a block diagonal matrix with each block $\boldsymbol{\Theta}_{g}$ being a complex symmetric unitary matrix, $\forall g$.


In the following, we propose two novel strategies to design discrete-value group and fully connected RISs.
In the first, scalar-discrete RISs, $B$ resolution bits are allocated to each reactance matrix entry $\left[\mathbf{X}_I\right]_{i,j}$.
In the second, vector-discrete RISs, $B_V$ bits are set to each reactance matrix block $\mathbf{X}_{I,g}$.

\section{Scalar-Discrete Group/Fully Connected RIS}
\label{sec:scalar-discrete}

Our goal is to design the discrete-value matrix $\boldsymbol{\Theta}$ and the vectors $\mathbf{g}$ and $\mathbf{w}$ to maximize the received signal power, given by $P_{R}=P_{T}\left|\mathbf{g}\left(\mathbf{H}_{RT}+\mathbf{H}_{RI}\boldsymbol{\Theta}\mathbf{H}_{IT}\right)\mathbf{w}\right|^{2}$.
In the case of single-stream transmission, the optimal precoding and combining vectors are given by the dominant eigenmode transmission \cite{cle13}.
Thus, maximizing the received signal power is equivalent to maximize $\left\|\mathbf{H}_{RT}+\mathbf{H}_{RI}\boldsymbol{\Theta}\mathbf{H}_{IT}\right\|^{2}$.
To investigate the design of discrete-value group and fully connected RISs, it is necessary to first consider the design of group and fully connected RISs with continuous-value $\boldsymbol{\Theta}$.
Then, the continuous-value optimization is used to optimize $\boldsymbol{\Theta}$ with discrete values.

\subsection{Continuous-Value Group/Fully Connected RIS}

The received signal power maximization problem for continuous-value $\boldsymbol{\Theta}$ in \gls{mimo} systems can be formulated as
\begin{align}
\underset{\mathbf{X}_{I,g}}{\mathsf{\mathrm{max}}}\;\;
& \left\|\mathbf{H}_{RT}+\mathbf{H}_{RI}\boldsymbol{\Theta}\mathbf{H}_{IT}\right\|^{2}\label{eq:P-GC-C-obj-0}\\
\mathsf{\mathrm{s.t.}}\;\;\; & \boldsymbol{\Theta}=\mathrm{diag}\left(\boldsymbol{\Theta}_{1},\ldots,\boldsymbol{\Theta}_{G}\right),\label{eq:P-GC-C-con1-0}\\
& \boldsymbol{\Theta}_{g}=\boldsymbol{\Theta}_{g}^{T},\:\boldsymbol{\Theta}_{g}^{H}\boldsymbol{\Theta}_{g}=\boldsymbol{\mathrm{I}},\:\forall g,\label{eq:P-GC-C-con2-0}
\end{align}
where the constraints derived from \eqref{eq:T group} indicate that the scattering matrix $\boldsymbol{\Theta}$ is a block diagonal matrix with each block being a complex symmetric unitary matrix.
These constraints complicate the optimization problem, which needs to be reformulated.
Thus, the relationship between $\boldsymbol{\Theta}$ and the reactance matrix $\mathbf{X}_{I}$ given by \eqref{eq:T(X)} is exploited to equivalently rewrite problem \eqref{eq:P-GC-C-obj-0}-\eqref{eq:P-GC-C-con2-0} as
\begin{align}
\underset{\mathbf{X}_{I,g}}{\mathsf{\mathrm{max}}}\;\;
& \left\|\mathbf{H}_{RT}+\mathbf{H}_{RI}\boldsymbol{\Theta}\mathbf{H}_{IT}\right\|^{2}\label{eq:P-GC-C-obj}\\
\mathsf{\mathrm{s.t.}}\;\;\; & \boldsymbol{\Theta}=\mathrm{diag}\left(\boldsymbol{\Theta}_{1},\ldots,\boldsymbol{\Theta}_{G}\right),\label{eq:P-GC-C-con1}\\
& \boldsymbol{\Theta}_{g}=\left(j\mathbf{X}_{I,g}+Z_{0}\mathbf{I}\right)^{-1}\left(j\mathbf{X}_{I,g}-Z_{0}\mathbf{I}\right),\:\forall g,\label{eq:P-GC-C-con2}\\
& \mathbf{X}_{I,g}=\mathbf{X}_{I,g}^{T},\:\forall g,\label{eq:P-GC-C-con3}
\end{align}
which can be transformed into an unconstrained problem.
More precisely, exploiting the constraints \eqref{eq:P-GC-C-con1} and \eqref{eq:P-GC-C-con2}, the objective $\left\|\mathbf{H}_{RT}+\mathbf{H}_{RI}\boldsymbol{\Theta}\mathbf{H}_{IT}\right\|^{2}$ can be expressed as a function of $\mathbf{X}_{I,g},\:\forall g$.
Since $\mathbf{X}_{I,g}$ is an arbitrary $N_{G}\times N_{G}$ real symmetric matrix, $\mathbf{X}_{I,g}$ is an unconstrained function of the $N_{G}\left(N_{G}+1\right)/2$ entries in its upper triangular part.
Thus, the obtained problem is an unconstrained optimization problem in the variables $\left[\mathbf{X}_{I,g}\right]_{i,j}$, with $i\leq j$ and $\forall g$.

Problem \eqref{eq:P-GC-C-obj}-\eqref{eq:P-GC-C-con3} has been solved for RIS-aided \gls{siso} systems in \cite{she20} by using the quasi-Newton method to find the optimal upper triangular part of each $\mathbf{X}_{I,g}$ without any constraints.
For RIS-aided \gls{mimo} systems, we propose to solve it by alternatively optimizing $\mathbf{X}_I$ and the beamforming vectors $\mathbf{g}$ and $\mathbf{w}$, as established in the literature on single connected RISs \cite{wu21}, \cite{wu19c}, \cite{zap21}.
After $\mathbf{g}$ and $\mathbf{w}$ are initialized to feasible values, this optimization process alternates between the two following steps until convergence is reached.
With fixed $\mathbf{g}$ and $\mathbf{w}$, we update $\mathbf{X}_I$ by maximizing the objective $\left|\mathbf{g}\mathbf{H}_{RT}\mathbf{w}+\mathbf{g}\mathbf{H}_{RI}\boldsymbol{\Theta}\mathbf{H}_{IT}\mathbf{w}\right|$ with the quasi-Newton method as in \cite{she20}, where $\boldsymbol{\Theta}$ is a function of $\mathbf{X}_I$ given by \eqref{eq:T(X)}.
With fixed $\mathbf{X}_I$, and consequently fixed $\boldsymbol{\Theta}$, we update $\mathbf{g}$ and $\mathbf{w}$ as the dominant left and right singular vectors of the matrix $\mathbf{H}_{RT}+\mathbf{H}_{RI}\boldsymbol{\Theta}\mathbf{H}_{IT}$, respectively.
Note that this method is proven to converge to a stationary point of the objective.

\subsection{Problem Formulation for Discrete-Value RIS}

In \eqref{eq:P-GC-C-obj}-\eqref{eq:P-GC-C-con3}, the reactance matrix entries are allowed to assume arbitrary real values, which is hard to realize in practice.
Thus, we are interested in an RIS design strategy in which each reactance matrix entry is selected from a codebook.
We consider a discrete reactance codebook $\mathcal{C}$ symmetric around zero, written as
\begin{equation}
\mathcal{C} = \left\{\pm c_{1},\pm c_{2},\ldots,\pm c_{2^{B-1}}\right\},\label{eq:codebook}
\end{equation}
where $c_{b}>0$, for $b=1,\ldots,2^{B-1}$.
When selecting each reactance matrix entry from a codebook, the RIS optimization problem cannot be simply solved as an unconstrained problem.
Thus, instead of \eqref{eq:P-GC-C-obj}-\eqref{eq:P-GC-C-con3}, the following optimization problem can be considered to optimize $\mathbf{X}_{I}$
\begin{align}
\underset{\mathbf{X}_{I,g}}{\mathsf{\mathrm{max}}}\;\;
& \left\|\mathbf{H}_{RT}+\mathbf{H}_{RI}\boldsymbol{\Theta}\mathbf{H}_{IT}\right\|^{2}\label{eq:P-GC-D-obj1}\\
\mathsf{\mathrm{s.t.}}\;\;\;
& \boldsymbol{\Theta}=\mathrm{diag}\left(\boldsymbol{\Theta}_{1},\ldots,\boldsymbol{\Theta}_{G}\right),\label{eq:P-GC-D-con1-1}\\
& \boldsymbol{\Theta}_{g}=\left(j\mathbf{X}_{I,g}+Z_{0}\mathbf{I}\right)^{-1}\left(j\mathbf{X}_{I,g}-Z_{0}\mathbf{I}\right),\:\forall g,\label{eq:P-GC-D-con1-2}\\
& \mathbf{X}_{I,g}=\mathbf{X}_{I,g}^{T},\:\forall g,\label{eq:P-GC-D-con1-3}\\
& \left[\mathbf{X}_{I,g}\right]_{i,j}\in\mathcal{C},\:\forall g.\label{eq:P-GC-D-con1-4}
\end{align}
Since the entries of $\mathbf{X}_{I}$ are not distributed in a finite interval, to define the optimal codebook $\mathcal{C}$ is not as straightforward as in the single connected case \cite{wu21}.
Thus, we introduce a second optimization problem to design $\mathcal{C}$ as
\begin{align}
\underset{\mathcal{C}}{\mathsf{\mathrm{max}}}\;\;
& \mathrm{E}\left[\left\|\mathbf{H}_{RT}+\mathbf{H}_{RI}\boldsymbol{\Theta}\mathbf{H}_{IT}\right\|^{2}\right]\label{eq:P-GC-D-obj2}\\
\mathsf{\mathrm{s.t.}}\;\;\;
& \boldsymbol{\Theta}=\mathrm{diag}\left(\boldsymbol{\Theta}_{1},\ldots,\boldsymbol{\Theta}_{G}\right),\label{eq:P-GC-D-con2-1}\\
& \boldsymbol{\Theta}_{g}=\left(j\mathbf{X}_{I,g}+Z_{0}\mathbf{I}\right)^{-1}\left(j\mathbf{X}_{I,g}-Z_{0}\mathbf{I}\right),\:\forall g,\label{eq:P-GC-D-con2-2}\\
& \mathbf{X}_{I,g}\text{ solves \eqref{eq:P-GC-D-obj1}-\eqref{eq:P-GC-D-con1-4}},\:\forall g,\label{eq:P-GC-D-con2-3}
\end{align}
where the objective is the ergodic received signal power and the expectation operator $\mathrm{E}[\cdot]$ represents the average over the channel realizations\footnote{Note that the received signal power in \eqref{eq:P-GC-D-obj1}-\eqref{eq:P-GC-D-con2-3} can be replaced by any objective function depending on the \gls{csi} and the RIS scattering matrix.
Thus, the proposed codebook design and optimization framework is general enough to accommodate any objective function, including the metrics typically considered in multi-user systems, e.g., the sum rate.}.
Note that in \eqref{eq:P-GC-D-obj2}-\eqref{eq:P-GC-D-con2-3}, $\mathbf{X}_{I,g}$ implicitly depends on $\mathcal{C}$ because of constraint \eqref{eq:P-GC-D-con2-3}.
These two nested optimization problems form a bilevel programming problem, in which \eqref{eq:P-GC-D-obj2}-\eqref{eq:P-GC-D-con2-3} is the upper-level problem, and \eqref{eq:P-GC-D-obj1}-\eqref{eq:P-GC-D-con1-4} is the lower-level problem.
This bilevel problem is difficult to solve since in the upper-level problem we do not have an expression for the constraint \eqref{eq:P-GC-D-con2-3}, and we cannot explicitly write $\mathbf{X}_{I,g}$ as a function of $\mathcal{C}$.

For this reason, when considering the optimization of discrete group and fully connected RISs, two problems arise.
Firstly, during the offline learning stage, a suitable discrete reactance codebook has to be defined for the elements of $\mathbf{X}_{I}$.
Secondly, during the online deployment stage, the discrete values of $\mathbf{X}_{I}$ are optimized by choosing among the values of the fixed codebook.
In the following, these two stages are described.
The offline learning stage is performed differently for binary-level ($B=1$) and multi-level ($B>1$) codebooks since a lower complexity solution is available when $B=1$.

\subsection{Offline Learning for Binary-Level Codebook}
\label{sec:scalar-discrete-binary}

During offline learning, we assume that the codebook design can exploit a training set of $N_0$ channel realization triplets.
Such a training set can be practically collected through a sampling campaign to be conducted offline.
During this offline sampling campaign, channel realizations are collected and stored with dedicated transceiver devices.
The $n_0$th triplet is denoted as $\mathcal{H}^{[n_0]}=(\mathbf{H}_{RT}^{[n_0]},\mathbf{H}_{RI}^{[n_0]},\mathbf{H}_{IT}^{[n_0]})$ and includes the channels from transmitter to receiver, RIS to receiver, and transmitter to RIS, respectively.
Thus, the training set is a set of triplets defined as
\begin{equation}
\mathcal{H}_0=\left\{\mathcal{H}^{[1]},\mathcal{H}^{[2]},\ldots,\mathcal{H}^{[N_0]}\right\}.
\end{equation}
When the number of resolution bits per reconfigurable reactance is $B=1$, the entries of the reactance matrix $\mathbf{X}_{I}$ can assume only the two values in $\mathcal{C}=\{-c_{1},+c_{1}\}$, where $c_{1}$ is a positive real number.
Exploiting this special structure of binary-level codebooks, we can obtain the optimal codebook value $c_{1}$ and binary-value reactance matrix $\bar{\mathbf{X}}_I$ for each channel triplet in the training set by solving the bilevel optimization problem \eqref{eq:P-GC-D-obj1}-\eqref{eq:P-GC-D-con2-3}.

Firstly, problem \eqref{eq:P-GC-D-obj1}-\eqref{eq:P-GC-D-con1-4} can be solved given a fixed value of $c_1$ through alternating optimization.
With this method, each non-zero entry of the reactance matrix is optimized individually by searching among the two possible values in $\mathcal{C}$, while fixing the other $N_I\left(N_{G}+1\right)/2-1$ entries.
This procedure is repeated for all the reactance matrix entries, iterating multiple times until the convergence is reached.
In this way, we can evaluate the optimal $\mathbf{X}_{I,g}$ as function of $c_{1}$ in constraint \eqref{eq:P-GC-D-con2-3}.
Secondly, the upper-level problem \eqref{eq:P-GC-D-obj2}-\eqref{eq:P-GC-D-con2-3} is solvable with one-dimensional pattern search since its objective can be readily evaluated as a function of $c_{1}$.
We employ pattern search since it is an optimization algorithm that only requires the evaluation of the objective function in a series of points until the convergence is reached, without using derivatives \cite{aud02}.
Thus, we build the two sets
\begin{align}
\mathcal{C}_1&=\left\{c_1^{\left[1\right]},c_1^{\left[2\right]},\ldots,c_{1}^{\left[N_0\right]}\right\},\\
\bar{\mathcal{X}}_I&=\left\{\bar{\mathbf{X}}_I^{\left[1\right]},\bar{\mathbf{X}}_I^{\left[2\right]},\ldots,\bar{\mathbf{X}}_I^{\left[N_0\right]}\right\},
\end{align}
where $c_{1}^{\left[n_0\right]}$ and $\bar{\mathbf{X}}_I^{\left[n_0\right]}$ are the optimal codebook value $c_{1}$ and binary-value reactance matrix $\bar{\mathbf{X}}_I$ associated to the channel triplet $\mathcal{H}^{[n_0]}$, respectively.

Given the two sets $\mathcal{C}_1$ and $\bar{\mathcal{X}}_I$, we design the optimal value for $c_1$, denoted as $c_1^\star$.
We obtain $c_{1}^\star$ as the value that minimizes the distortion caused in the reactance matrix
\begin{align}
c_1^\star=\argmin_{c_{1}}\;\;
& \frac{1}{N_0}\sum_{n_0=1}^{N_0}\left\Vert\bar{\mathbf{X}}_I^{[n_0]}-\mathbf{X}_I\right\Vert_F^2\label{eq:B1-1}\\
\mathsf{\mathrm{s.t.}}\;\;\;
& \mathbf{X}_{I}=\mathrm{diag}\left(\mathbf{X}_{I,1},\ldots,\mathbf{X}_{I,G}\right),\\
& \left[\mathbf{X}_{I,g}\right]_{i,j}\in\left\{-c_1,+c_1\right\},\:\forall g,
\end{align}
where \eqref{eq:B1-1} is a sample average computed over the channel triplets in the training set.
In \eqref{eq:B1-1}, each non-zero squared entry of the matrix $\bar{\mathbf{X}}_I^{[n_0]}-\mathbf{X}_I$ is given by $(c_1^{[n_0]}-c_1)^2$ because of the symmetry of the codebook.
Thus, problem \eqref{eq:B1-1} can be reformulated as
\begin{equation}
c_1^\star=\argmin_{c_{1}}\;\;\frac{N_IN_G}{N_0}\sum_{n_0=1}^{N_0}\left(c_1^{[n_0]}-c_1\right)^2.\label{eq:B1-2}
\end{equation}
By setting the derivative of the objective function in \eqref{eq:B1-2} with respect to $c_1$ to zero, we obtain
\begin{equation}
c_1^\star=\frac{1}{N_0}\sum_{n_0=1}^{N_0}c_1^{[n_0]}.\label{eq:B1-3}
\end{equation}
Finally, the binary-value codebook is given by $\mathcal{C}=\{-c_1^\star,+c_1^\star\}$.

The complexity of the offline learning for binary-level codebook is driven by the complexity of solving the bilevel problem \eqref{eq:P-GC-D-obj1}-\eqref{eq:P-GC-D-con2-3} $N_0$ times to find $\mathcal{C}_1$.
Since one-dimensional pattern search requires computing twice the objective function per iteration, the complexity per iteration is given by $\mathcal{O}(2N_0N_I(N_G+1))$.

\subsection{Offline Learning for Multi-Level Codebook}
\label{sec:scalar-discrete-multi}

Pattern search can be used to solve the upper-level problem \eqref{eq:P-GC-D-obj2}-\eqref{eq:P-GC-D-con2-3} only when the optimization involves a one-dimensional search, that is only when $B=1$.
When more resolution bits are considered, pattern search becomes highly sub-optimal since it is likely to converge to a local maximum in a multi-dimensional search space.
For this reason, we rely on a different method to define multi-level codebooks.
The main idea is to learn the codebook from the distribution of the optimal continuous-value $\mathbf{X}_{I}$ which solve \eqref{eq:P-GC-C-obj}-\eqref{eq:P-GC-C-con3}.
More precisely, we find the optimal set of values $\{c_1^\star,\ldots,c_K^\star\}$, with $K=2^{B-1}$, such that the distortion on the optimal continuous-value reactance matrix is minimized.
The process leading to the multi-level codebook $\mathcal{C}=\{\pm c_1^\star,\ldots,\pm c_{2^{B-1}}^\star\}$ is detailed in the following.

We firstly solve \eqref{eq:P-GC-C-obj}-\eqref{eq:P-GC-C-con3} for each training set triplet by alternatively optimizing $\mathbf{X}_I$ and the beamforming vectors $\mathbf{g}$ and $\mathbf{w}$.
As a result, we construct the set
\begin{equation}
\mathcal{X}_I=\left\{\mathbf{X}_I^{\left[1\right]},\mathbf{X}_I^{\left[2\right]},\ldots,\mathbf{X}_I^{\left[N_0\right]}\right\},
\end{equation}
where $\mathbf{X}_I^{\left[n_0\right]}$ is the optimal continuous-value reactance matrix associated to the channel triplet $\mathcal{H}^{[n_0]}$.
Given the set $\mathcal{X}_I$, the optimal values $\{c_1^\star,\ldots,c_K^\star\}$ are determined by solving
\begin{align}
\{c_1^\star,\ldots,c_K^\star\}=\argmin_{\{c_1,\ldots,c_K\}}\;\;
& \frac{1}{N_0}\sum_{n_0=1}^{N_0}\left\Vert\mathbf{X}_I^{[n_0]}-\mathbf{X}_I\right\Vert_F^2\label{eq:B2-1}\\
\mathsf{\mathrm{s.t.}}\;\;\;
& \mathbf{X}_{I}=\mathrm{diag}\left(\mathbf{X}_{I,1},\ldots,\mathbf{X}_{I,G}\right),\\
& \left[\mathbf{X}_{I,g}\right]_{i,j}\in\left\{\pm c_1,\ldots,\pm c_K\right\},\:\forall g.
\end{align}
To simplify problem \eqref{eq:B2-1}, we introduce the vectors $\mathbf{x}^{[n_0]}\in\mathbb{R}^{1\times N_IN_G}$ containing the non-zero entries of $\mathbf{X}_I^{[n_0]}$.
Then, the $N_0$ vectors $\mathbf{x}^{[n_0]}$ are concatenated in $\mathbf{x}_0\in\mathbb{R}^{1\times N_0N_IN_G}$ as $\mathbf{x}_0=[\mathbf{x}^{[1]},\ldots,\mathbf{x}^{[N_0]}]$.
In this way, problem \eqref{eq:B2-1} becomes
\begin{align}
\{c_1^\star,\ldots,c_K^\star\}=\argmin_{\{c_1,\ldots,c_K\}}\;\;
& \frac{1}{N_0}\sum_{n=1}^{N_0N_IN_G}\left\vert\left[\mathbf{x}_0\right]_n-\left[\mathbf{x}_I\right]_n\right\vert^2\label{eq:B2-2}\\
\mathsf{\mathrm{s.t.}}\;\;\;
& \left[\mathbf{x}_I\right]_n\in\left\{\pm c_1,\ldots,\pm c_K\right\},\:\forall n.
\end{align}
where the vector $\mathbf{x}_I\in\mathbb{R}^{1\times N_0N_IN_G}$ has been introduced as an auxiliary variable.
Exploiting the symmetry of the codebook $\mathcal{C}$, \eqref{eq:B2-2} can be further reformulated as
\begin{align}
\{c_1^\star,\ldots,c_K^\star\}=\argmin_{\{c_1,\ldots,c_K\}}\;\;
& \frac{1}{N_0}\sum_{n=1}^{N_0N_IN_G}\left\vert\left\vert\left[\mathbf{x}_0\right]_n\right\vert-\left[\mathbf{x}_I\right]_n\right\vert^2\label{eq:B2-3}\\
\mathsf{\mathrm{s.t.}}\;\;\;
& \left[\mathbf{x}_I\right]_n\in\left\{c_1,\ldots,c_K\right\},\:\forall n,
\end{align}
which is a $K$-means clustering problem in a one-dimensional space, with $K=2^{B-1}$ \cite{bis06}.
For each data point $x_n=\left\vert\left[\mathbf{x}_0\right]_n\right\vert$ we introduce the binary indicator $r_{nk}\in\{0,1\}$ such that $r_{nk}=1$ if $x_n$ is assigned to $c_k$, and $r_{ni}=0$ otherwise.
The resulting $K$-means clustering problem is formalized as
\begin{equation}
\{c_1^\star,\ldots,c_K^\star\}=\argmin_{\{c_1,\ldots,c_K\}}\;\;\sum_{n=1}^{N_0N_IN_G}\sum_{k=1}^{2^{B-1}}r_{nk}\left\vert x_n-c_k\right\vert^2\label{eq:B2-4}
\end{equation}
To solve \eqref{eq:B2-4}, we iteratively optimize the sets $\{r_{nk}|n=1,\ldots,N_0N_IN_G,k=1,\ldots,K\}$ and $\{c_1,\ldots,c_K\}$.
Firstly, when the values $\{c_1,\ldots,c_K\}$ are fixed, we set $\{r_{nk}\}$ by assigning the $n$th data point $x_n$ to the closest cluster, i.e.,
\begin{equation}
r_{nk}=
\begin{cases}
1 & \text{if } k=\arg \min_{j} \left\vert x_n-c_j\right\vert^2\\
0 & \text{otherwise}
\end{cases},
\end{equation}
for $n=1,\ldots,N_0N_IN_G$ and $k=1,\ldots,K$.
Secondly, when $\{r_{nk}\}$ are fixed, the cluster centers $\{c_1,\ldots,c_K\}$ are updated.
This can be done in close form by setting the derivative of the objective function in \eqref{eq:B2-4} with respect to $c_k$ to zero
\begin{equation}
c_k=\frac{\sum_n r_{nk}x_n}{\sum_n r_{nk}},
\end{equation}
for $k=1,\ldots,K$.
The two steps of assigning data points to clusters and updating the cluster
centers are repeated alternatively until there is no further change in the cluster assignment.
Finally, the multi-value codebook is given by $\mathcal{C}=\{\pm c_1^\star,\ldots,\pm c_{2^{B-1}}^\star\}$.

The complexity of the offline learning for multi-level codebook is given by $\mathcal{O}(N_0(N_I(N_G+1)/2)^2+N_0N_IN_G2^{B-1})$, where the first term is due to solving \eqref{eq:P-GC-C-obj}-\eqref{eq:P-GC-C-con3} $N_0$ times to find $\mathcal{X}_I$, and the second term is due to the $K$-means clustering algorithm used to solve \eqref{eq:B2-4}.

\begin{algorithm}[t]
\KwIn{$B$, $\mathcal{H}_0$, $\mathbf{H}_{RT}$, $\mathbf{H}_{RI}$, $\mathbf{H}_{IT}$, $N_G$}
\KwOut{$\mathcal{C}$, $\mathbf{X}_{I}$}
\nonl Offline Learning\\
\uIf{$B==1$}{
Calculate $\mathcal{C}_1$ by the bilevel problem \eqref{eq:P-GC-D-obj1}-\eqref{eq:P-GC-D-con2-3}\;
Calculate $c_1^\star$ by \eqref{eq:B1-3}\;
$\mathcal{C}\leftarrow\{-c_1^\star,+c_1^\star\}$\;
}
\Else{
Calculate $\mathcal{X}_I$ by \eqref{eq:P-GC-C-obj}-\eqref{eq:P-GC-C-con3}\;
Calculate $\{c_1^\star,\ldots,c_K^\star\}$ by iteratively solving \eqref{eq:B2-4}, with $K=2^{B-1}$\;
$\mathcal{C}\leftarrow\{\pm c_1^\star,\ldots,\pm c_{K}^\star\}$\;
}
\nonl Online Deployment\\
Initialize $\mathbf{X}_{I,g}=\mathbf{0},\:\forall g$\;
\While{no convergence of objective \eqref{eq:P-GC-D-obj1}}{
\For{$g\leftarrow 1$ \KwTo $G$}{
\For{$i\leftarrow 1$ \KwTo $N_G$}{
\For{$j\leftarrow i$ \KwTo $N_G$}{
Update $\left[\mathbf{X}_{I,g}\right]_{i,j}\in\mathcal{C}$ by searching among the $2^B$ possible values\;
$\left[\mathbf{X}_{I,g}\right]_{j,i}\leftarrow\left[\mathbf{X}_{I,g}\right]_{i,j}$
}
}
}
}
\caption{Scalar-Discrete RISs Design}
\label{alg:scalar}
\end{algorithm}

\subsection{Online Deployment}


Once the codebook has been learned offline, the discrete $\mathbf{X}_{I}$ entries are optimized during the online deployment stage.
This is achieved by solving \eqref{eq:P-GC-D-obj1}-\eqref{eq:P-GC-D-con1-4} depending on the instantaneous \gls{csi} $\mathbf{H}_{RT}$, $\mathbf{H}_{RI}$, and $\mathbf{H}_{IT}$ and on the designed codebook $\mathcal{C}$.
To solve this problem, an exhaustive search could be employed since each reactance matrix entry can assume a finite number of values.
However, to reduce the search space, we employ an alternating optimization method.
In this sub-optimal method, each non-zero entry of the reactance matrix is optimized individually by searching among all possible $2^{B}$ values, while fixing the other $N_I\left(N_{G}+1\right)/2-1$ entries.
This procedure is repeated for all $N_I\left(N_{G}+1\right)/2$ reactance matrix entries, iterating multiple times until the convergence is reached.
The convergence is considered reached when the fractional increase of the objective value in a full iteration is below a certain parameter $\epsilon$.
The convergence is guaranteed by the following two facts.
First, at each iteration, the objective function $\Vert\mathbf{H}_{RT}+\mathbf{H}_{RI}\boldsymbol{\Theta}\mathbf{H}_{IT}\Vert^{2}$ is non-decreasing.
This holds since each reactance matrix entry is optimized by exhaustively searching among the possible values in the codebook $\mathcal{C}$.
Second, the objective function is bounded from above by $(\Vert\mathbf{H}_{RT}\Vert+\Vert\mathbf{H}_{RI}\Vert\Vert\mathbf{H}_{IT}\Vert)^2$.
The design of scalar-discrete RISs is summarized in Alg.~\ref{alg:scalar}, which solves the problem \eqref{eq:P-GC-D-obj1}-\eqref{eq:P-GC-D-con2-3}.

Note that we assume that the statistical properties of the channels are unchanged during the offline learning and online deployment stages.
This is necessary since we approximate the expectation of the received signal power with its sample average, computed over the training set.
However, numerical simulations showed that learning the codebook based on outdated channel statistics only slightly degrades the performance.
Besides, perfect instantaneous \gls{csi} is assumed during the online deployment stage, which can be obtained with the semi-passive channel estimation strategy proposed in \cite{wu21}.

\subsection{Scaling Law}

To analyze the optimality of the codebook in \eqref{eq:codebook}, we characterize the scaling law of the received signal power as a function of the number of RIS elements $N_I$, when $\mathcal{C}$ is used to discretize the reactance matrix.
Note that this analysis is necessary since the scaling law of the proposed discrete RIS architectures is not known and not straightforward given the novel constraints of BD-RISs.
To obtain fundamental insights, we consider a \gls{siso} system, i.e., $N_T=1$ and $N_R=1$.
Furthermore, the direct link $h_{RT}$ is negligible in comparison with the reflected signal power for asymptotically large $N_I$.
Considering unitary transmit power, the received signal power is given by $P_R=\left|\mathbf{h}_{RI}\boldsymbol{\Theta}\mathbf{h}_{IT}\right|^2$, where $\boldsymbol{\Theta}$ is obtained by discretizing the reactance matrix $\mathbf{X}_I$.
Such a scaling law is given in the following proposition.
\begin{proposition}
Assume i.i.d. Rayleigh fading channels, i.e., $\mathbf{h}_{RI}\sim\mathcal{CN}\left(\mathbf{0},\mathbf{I}\right)$ and $\mathbf{h}_{IT}\sim\mathcal{CN}\left(\mathbf{0},\mathbf{I}\right)$.
A group connected RIS whose reactance matrix entries are chosen from $\mathcal{C}$ can achieve an average received signal power in the order of $\mathcal{O}\left(N_I^2\right)$ as $N_I$ increases, even with one resolution bit per each reactance.
\label{pro:1}
\end{proposition}
\begin{proof}
Let us consider the worst case $B=1$, and assume that the matrix block $\mathbf{X}_{I,g}$ can only assume two values $\mathbf{X}_{I,g}=\pm c_1\text{ones}\left(N_G\right)$.
Note that proving Proposition 1 for the case $\mathcal{C}=\{\pm c_1\}$ also establishes its validity for all codebooks $\mathcal{C}=\{\pm c_{1},\pm c_{2},\ldots,\pm c_{2^{B-1}}\}$.
Through eigenmode decomposition, we write $\mathbf{X}_{I,g}=\mathbf{V}_{g}\boldsymbol{\Lambda}_{g}\mathbf{V}_{g}^T$, where $\boldsymbol{\Lambda}_{g}=\text{diag}\left(\boldsymbol{\lambda}_g\right)$ is diagonal containing the eigenvalues of $\mathbf{X}_{I,g}$ and $\mathbf{V}_{g}$ is orthonormal.
Among the eigenvalues of $\mathbf{X}_{I,g}$, only $[\boldsymbol{\lambda}_g]_1$ is non zero since $\mathbf{X}_{I,g}$ is rank one.
By applying \eqref{eq:T(X)}, the scattering matrix $\boldsymbol{\Theta}_g$ can be expressed as $\boldsymbol{\Theta}_g=\mathbf{V}_{g}\mathbf{D}_g\mathbf{V}_{g}^T$,
where $\mathbf{D}_g=\text{diag}\left(\mathbf{d}_g\right)$ is a diagonal matrix with $[\mathbf{d}_g]_{n_G}=\frac{j[\boldsymbol{\lambda}_g]_{n_{G}}-Z_{0}}{j[\boldsymbol{\lambda}_g]_{n_G}+Z_{0}}$.
According to \cite{she20}, the average received signal power for group connected RISs is written as
\begin{align}
\text{E}\left[P_{R}\right]
& =\text{E}\left[\left|\sum_{g=1}^{G}\mathbf{h}_{RI,g}\boldsymbol{\Theta}_{g}\mathbf{h}_{IT,g}\right|^2\right]\\
& =\text{E}\left[\left|\sum_{g=1}^{G}\sum_{n_G=1}^{N_G}\left[\bar{\mathbf{h}}_{RI,g}\right]_{n_G}\left[\mathbf{d}_g\right]_{n_G}\left[\bar{\mathbf{h}}_{IT,g}\right]_{n_G}\right|^2\right],
\end{align}
where we considered $\boldsymbol{\Theta}_g=\mathbf{V}_{g}\mathbf{D}_g\mathbf{V}_{g}^T$ and introduced  $\bar{\mathbf{h}}_{RI,g}=\mathbf{h}_{RI,g}\mathbf{V}_{g}$ and $\bar{\mathbf{h}}_{IT,g}=\mathbf{V}_{g}^T\mathbf{h}_{IT,g}$.
Recalling that $[\mathbf{d}_g]_{n_G}=-1$ if $n_G\geq 2$, we have
\begin{align}
\text{E}\left[P_{R}\right]
& =\text{E}\left[\left|\left(\sum_{g=1}^{G}\left[\bar{\mathbf{h}}_{RI,g}\right]_{1}\left[\mathbf{d}_g\right]_{1}\left[\bar{\mathbf{h}}_{IT,g}\right]_{1}\right)\right.\right.\\
& -\left.\left.\left(\sum_{g=1}^{G}\sum_{n_G=2}^{N_G}\left[\bar{\mathbf{h}}_{RI,g}\right]_{n_G}\left[\bar{\mathbf{h}}_{IT,g}\right]_{n_G}\right)\right|^2\right]\\
& >\text{E}\left[\left|\sum_{g=1}^{G}\left[\bar{\mathbf{h}}_{RI,g}\right]_1\left[\mathbf{d}_g\right]_1\left[\bar{\mathbf{h}}_{IT,g}\right]_1\right|^2\right]\\
& =\text{E}\left[\left|\sum_{g=1}^{G}\left|\left[\bar{\mathbf{h}}_{RI,g}\right]_1\right|\left|\left[\bar{\mathbf{h}}_{IT,g}\right]_1\right|e^{j\left[\text{arg}\left(\left[\mathbf{d}_g\right]_1\right)-\theta_g^\star\right]}\right|^2\right],\label{eq:last-P}
\end{align}
where we introduced $\theta_g^\star=-\text{arg}\left([\bar{\mathbf{h}}_{RI,g}]_1\right)-\text{arg}\left([\bar{\mathbf{h}}_{IT,g}]_1\right)$.
We denote as $\theta_1=\arg\left(\frac{j[\boldsymbol{\lambda}_g^+]_1-Z_{0}}{j[\boldsymbol{\lambda}_g^+]_1+Z_{0}}\right)\in\left(0,\pi\right)$, where $[\boldsymbol{\lambda}_g^+]_1$ is the non zero eigenvalue of $\mathbf{X}_{I,g}^+=+c_1\text{ones}\left(N_G\right)$.
Thus, we have $-\theta_1=\arg\left(\frac{j[\boldsymbol{\lambda}_g^-]_1-Z_{0}}{j[\boldsymbol{\lambda}_g^-]_1+Z_{0}}\right)\in\left(-\pi,0\right)$, where $[\boldsymbol{\lambda}_g^-]_1=-[\boldsymbol{\lambda}_g^+]_1$ is the non zero eigenvalue of $\mathbf{X}_{I,g}^-=-c_1\text{ones}\left(N_G\right)$.
We consider a quantization approach giving $\text{arg}\left([\mathbf{d}_g]_1\right)=\theta_1$ if $\theta_g^\star\in\left[0,\pi\right)$, or $\text{arg}\left([\mathbf{d}_g]_1\right)=-\theta_1$ otherwise.
The quantization error is denoted as $\bar{\theta}_g=\text{arg}\left([\mathbf{d}_g]_1\right)-\theta_g^\star$.
Thus, \eqref{eq:last-P} can be rewritten as
\begin{multline}
\text{E}\left[P_{R}\right]
>\text{E}\Bigg[\sum_{g=1}^{G}\left|\left[\bar{\mathbf{h}}_{RI,g}\right]_1\right|^2\left|\left[\bar{\mathbf{h}}_{IT,g}\right]_1\right|^2\\
+\sum_{g\neq f}\left|\left[\bar{\mathbf{h}}_{RI,g}\right]_1\right|\left|\left[\bar{\mathbf{h}}_{IT,g}\right]_1\right|\left|\left[\bar{\mathbf{h}}_{RI,f}\right]_1\right|\left|\left[\bar{\mathbf{h}}_{IT,f}\right]_1\right|e^{j\bar{\theta}_g-j\bar{\theta}_f}\Bigg].
\end{multline}
Noting that $\left|[\bar{\mathbf{h}}_{RI,g}]_1\right|$, $\left|[\bar{\mathbf{h}}_{IT,g}]_1\right|$, and $e^{j\bar{\theta}_g}$ are independent with each other, with $\text{E}\left[\left|[\bar{\mathbf{h}}_{RI,g}]_1\right|^2\right]=1$, $\text{E}\left[\left|[\bar{\mathbf{h}}_{RI,g}]_1\right|\right]=\sqrt{\pi}/2$, and $\text{E}\left[e^{j\bar{\theta}_g}\right]=\text{E}\left[e^{-j\bar{\theta}_g}\right]=\frac{2}{\pi}\sin\left(\theta_1\right)$, we have
\begin{align}
\text{E}\left[P_{R}\right]
& >G+G\left(G-1\right)\left(\frac{\sqrt{\pi}}{2}\right)^4\left(\frac{2}{\pi}\sin\left(\theta_1\right)\right)^2\\
& >N_I^2\left(\frac{\sin\left(\theta_1\right)}{2N_G^2}\right)^2.\label{eq:O(N^2)}
\end{align}
Since $\sin\left(\theta_1\right)>0$, \eqref{eq:O(N^2)} proves that $\mathcal{C}$ can achieve an average received signal power growth of $\mathcal{O}\left(N_I^2\right)$ as $N_I$ increases.
\end{proof}

According to Proposition~\ref{pro:1}, the entries of an RIS reactance matrix can be chosen from a codebook $\mathcal{C}$ with no degradation in the received signal power growth.
This justifies the use of a symmetric codebook for our discrete-value design strategy.

\section{Vector-Discrete Group/Fully Connected RIS}
\label{sec:vector-discrete}

In this section, we propose a second discretization strategy based on vector quantization.
To realize vector-discrete RISs, we assign $B_V$ resolution bits to each reactance matrix block $\mathbf{X}_{I,g}$.
Thus, we introduce the codebook $\mathcal{C}_V$ as a set of vectors
\begin{equation}
\mathcal{C}_{V} = \left\{\mathbf{c}_{1},\mathbf{c}_{2},\ldots,\mathbf{c}_{2^{B_V}}\right\}.
\end{equation}
where $\mathbf{c}_{k}\in\mathbb{R}^{\frac{N_{G}\left(N_{G}+1\right)}{2}}$, for $k=1,\ldots,2^{B_V}$.
Here, $\mathbf{c}_{k}$ is a possible value that the upper triangular part of each block $\mathbf{X}_{I,g}$ can assume.

\subsection{Offline Learning}

Similarly to the scalar discretization case, $\mathcal{C}_V$ can be obtained with $K$-means clustering through an offline learning phase, with the difference that a multi-dimensional feature space with dimension $N_{G}\left(N_{G}+1\right)/2$ is now considered, and $K=2^{B_{V}}$.
%
To construct the codebook $\mathcal{C}_V$, let us consider the set $\mathcal{X}_I$ containing the optimal reactance matrices associated with the $N_0$ training channel triplets.
To formalize the $K$-means clustering problem, we introduce the matrices $\mathbf{X}^{[n_0]}\in\mathbb{R}^{N_{G}\left(N_{G}+1\right)/2\times G}$ containing in their $g$th column the entries of the upper triangular part of $\mathbf{X}_{I,g}^{[n_0]}$, for $g=1,\ldots,G$.
Then, the $N_0$ matrices $\mathbf{X}^{[n_0]}$ are concatenated in $\mathbf{X}_0\in\mathbb{R}^{N_{G}\left(N_{G}+1\right)/2\times N_0G}$ as $\mathbf{X}_0=[\mathbf{X}^{[1]},\ldots,\mathbf{X}^{[N_0]}]$.
In this way, the resulting $K$-means clustering problem writes as
\begin{equation}
\{\mathbf{c}_1^\star,\ldots,\mathbf{c}_K^\star\}=\argmin_{\{\mathbf{c}_1,\ldots,\mathbf{c}_K\}}\;\;\sum_{n=1}^{N_0G}\sum_{k=1}^{2^{B_V}}r_{nk}\left\vert \mathbf{x}_n-\mathbf{c}_k\right\vert^2,\label{eq:BV}
\end{equation}
where the $n$th data point $\mathbf{x}_n\in\mathbb{R}^{N_{G}\left(N_{G}+1\right)/2\times 1}$ is the $n$th column of the matrix $\mathbf{X}_0$.
Remarkably, \eqref{eq:BV} is in the form of a $K$-means clustering problem.
Thus, we solve \eqref{eq:BV} by alternatively optimizing the sets $\{r_{nk}\}$ and $\{\mathbf{c}_1,\ldots,\mathbf{c}_K\}$, as described in the following.
Firstly, with fixed $\{\mathbf{c}_1,\ldots,\mathbf{c}_K\}$ are fixed, we set $\{r_{nk}\}$ by assigning the $n$th data point $\mathbf{x}_n$ to the closest cluster center $\mathbf{c}_k$, yielding
\begin{equation}
r_{nk}=
\begin{cases}
1 & \text{if } k=\arg \min_{j}\left\Vert \mathbf{x}_n-\mathbf{c}_j\right\Vert^2\\
0 & \text{otherwise}
\end{cases},
\end{equation}
for $n=1,\ldots,N_0G$ and $k=1,\ldots,K$.
Secondly, with fixed $\{r_{nk}\}$, the cluster centers $\{\mathbf{c}_1,\ldots,\mathbf{c}_K\}$ are optimally updated in closed-form as
\begin{equation}
\mathbf{c}_k=\frac{\sum_n r_{nk}\mathbf{x}_n}{\sum_n r_{nk}},
\end{equation}
for $k=1,\ldots,K$.
These two steps are alternatively repeated until convergence.
Finally, the vector-quantization codebook is given by $\mathcal{C}_V=\{\mathbf{c}_1^\star,\ldots,\mathbf{c}_{K}^\star\}$.

The complexity of the offline learning for vector-discrete group and fully connected RISs is given by $\mathcal{O}(N_0(N_I(N_G+1)/2)^2+N_0N_I(N_G+1)2^{B_V-1})$, where the first term is due to solving \eqref{eq:P-GC-C-obj}-\eqref{eq:P-GC-C-con3} $N_0$ times to find $\mathcal{X}_I$, and the second term is due to the $K$-means clustering algorithm used to solve \eqref{eq:BV}.

\begin{figure*}[t]
    \begin{centering} 
    \includegraphics[width=0.24\textwidth]{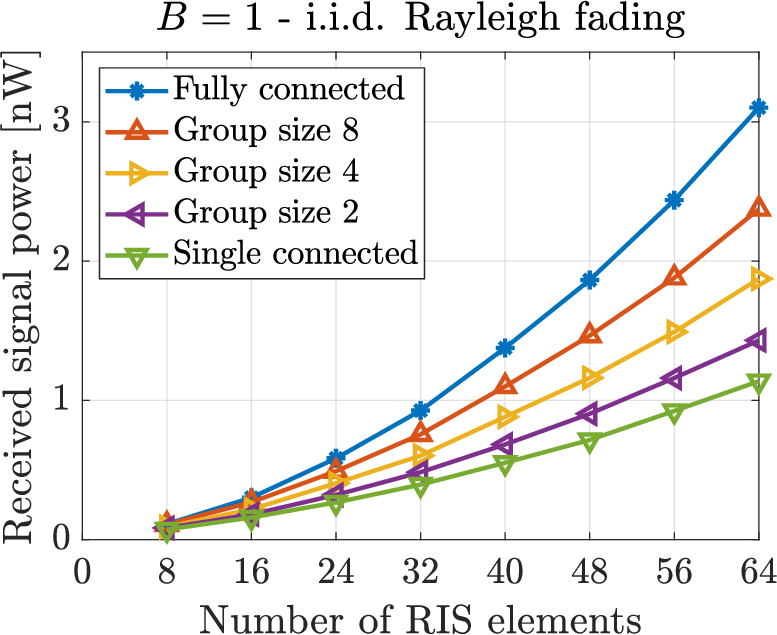}
    \includegraphics[width=0.24\textwidth]{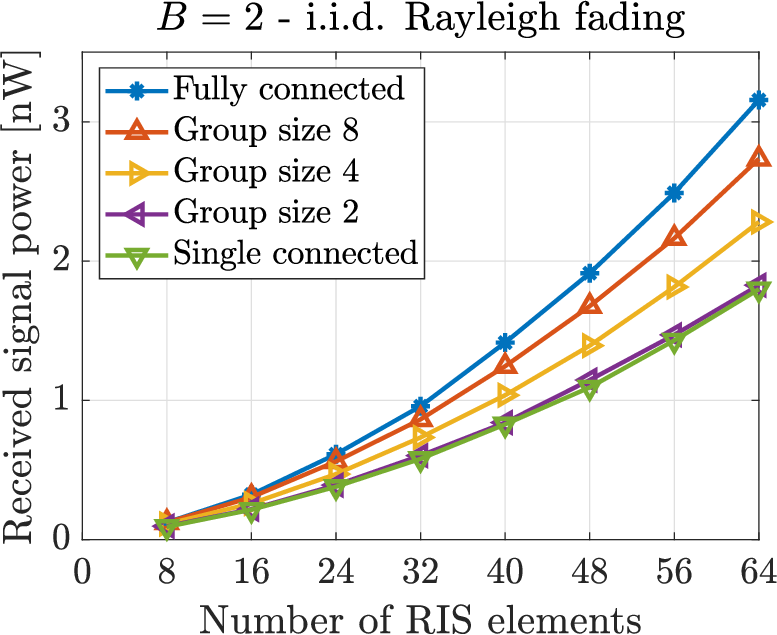}
    \includegraphics[width=0.24\textwidth]{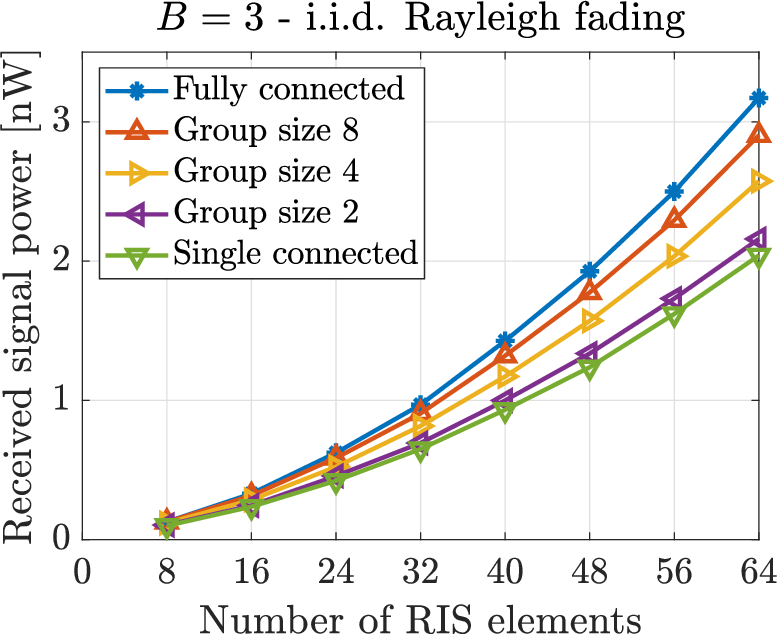}
    \includegraphics[width=0.24\textwidth]{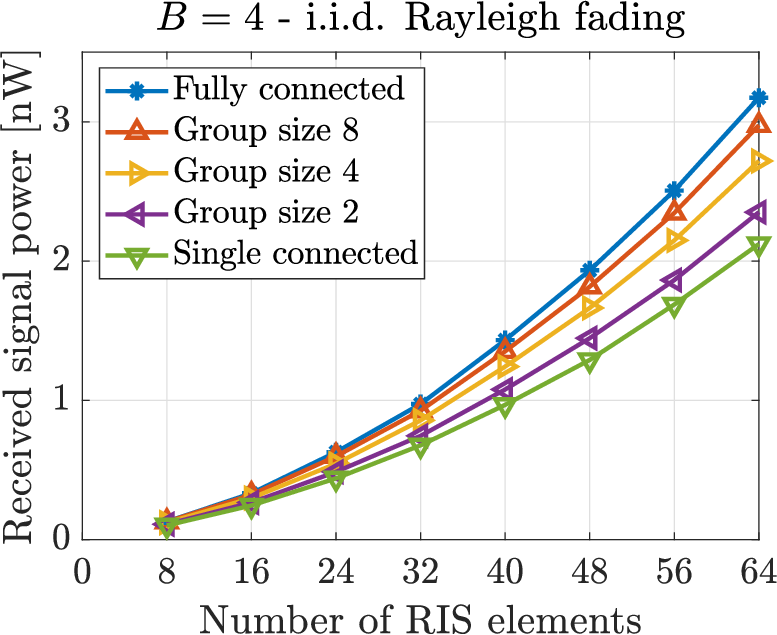}
    \par\end{centering}
    \vspace{0.1cm}
    \begin{centering}
    \includegraphics[width=0.24\textwidth]{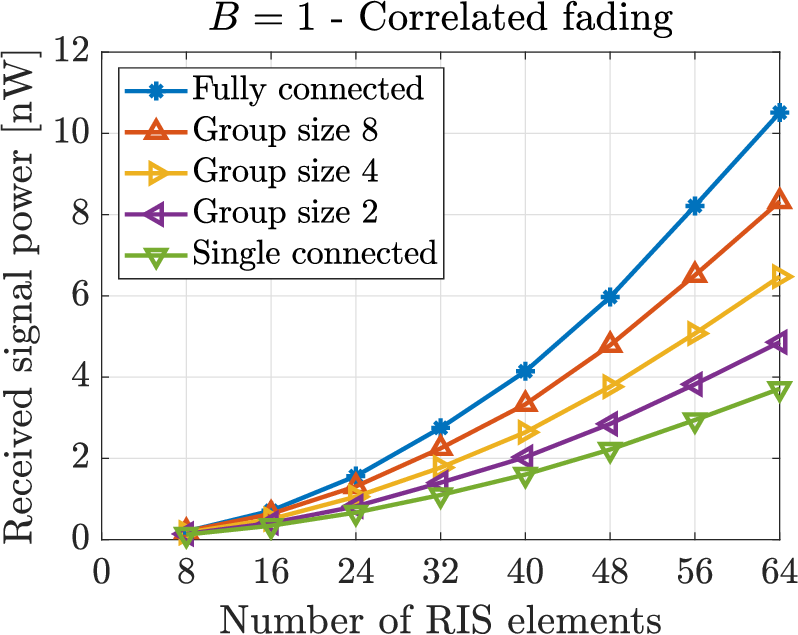}
    \includegraphics[width=0.24\textwidth]{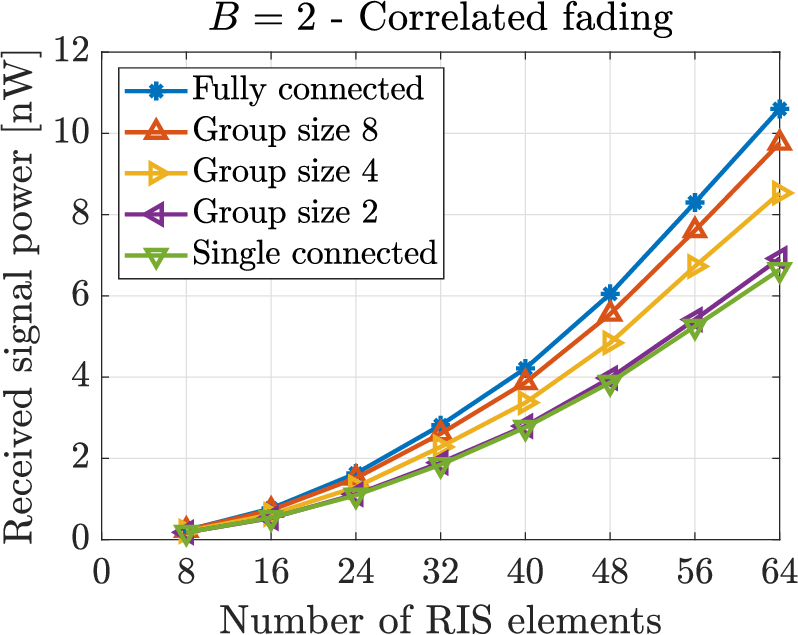}
    \includegraphics[width=0.24\textwidth]{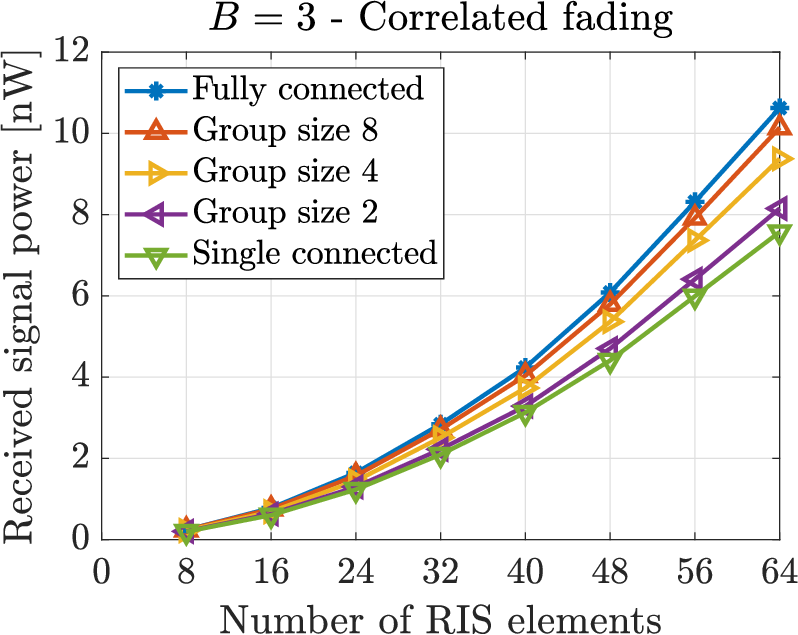}
    \includegraphics[width=0.24\textwidth]{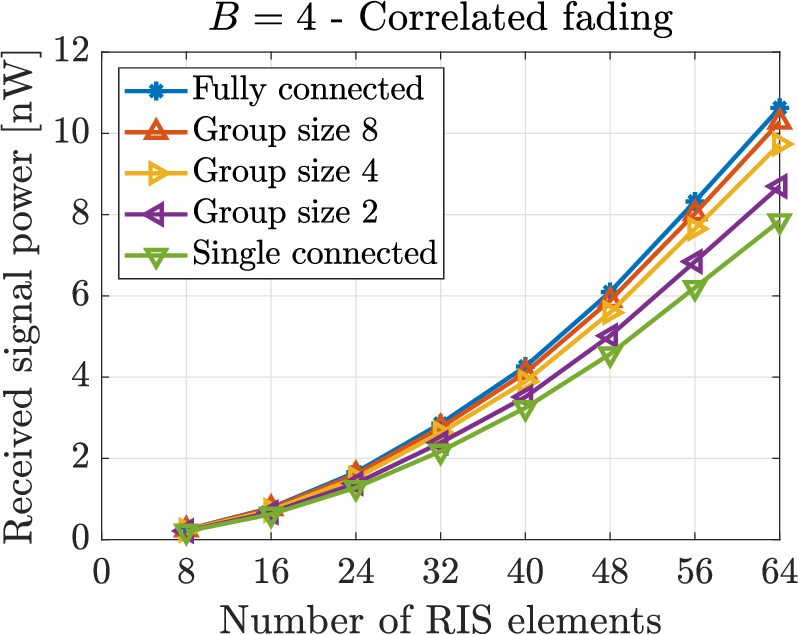}
    \par\end{centering}
    \caption{Average received signal power versus the number of RIS elements for different values of $B$.}
    \label{fig:bit}
\end{figure*}

\subsection{Online Deployment}

Once the codebook has been learned offline, the online optimization problem is formulated as
\begin{align}
\underset{\mathbf{X}_{I,g}}{\mathsf{\mathrm{max}}}\;\;
& \left\|\mathbf{H}_{RT}+\mathbf{H}_{RI}\boldsymbol{\Theta}\mathbf{H}_{IT}\right\|^{2}\label{eq:P-GC-D-obj4}\\
\mathsf{\mathrm{s.t.}}\;\;\;
& \boldsymbol{\Theta}=\mathrm{diag}\left(\boldsymbol{\Theta}_{1},\ldots,\boldsymbol{\Theta}_{G}\right),\label{eq:P-GC-D-con4-1}\\
& \boldsymbol{\Theta}_{g}=\left(j\mathbf{X}_{I,g}+Z_{0}\mathbf{I}\right)^{-1}\left(j\mathbf{X}_{I,g}-Z_{0}\mathbf{I}\right),\:\forall g,\label{eq:P-GC-D-con4-2}\\
& \mathbf{X}_{I,g}=\mathbf{X}_{I,g}^{T},\:\forall g,\label{eq:P-GC-D-con4-3}\\
& \mathbf{X}_{I,g}\in\left\{\mathbf{c}_1^\star,\ldots,\mathbf{c}_{2^{B_V}}^\star\right\},\:\forall g.\label{eq:P-GC-D-con4-4}
\end{align}
Alternating optimization is the selected strategy to solve this problem, in which the $G$ blocks $\mathbf{X}_{I,g}$ are iteratively optimized by searching among their possible $2^{B_V}$ values contained in $\mathcal{C}_V$.
The convergence of this iterative process can be proved as done for scalar-discrete RISs in Section~III~E.
The proposed design of vector-discrete RISs is summarized in Alg.~\ref{alg:vector}, where problem \eqref{eq:BV} is solved offline to design the codebook while \eqref{eq:P-GC-D-obj4}-\eqref{eq:P-GC-D-con4-4} is solved online to optimize the RIS reactance matrix.

\begin{algorithm}[t]
\KwIn{$B_V$, $\mathcal{H}_0$, $\mathbf{H}_{RT}$, $\mathbf{H}_{RI}$, $\mathbf{H}_{IT}$, $N_G$}
\KwOut{$\mathcal{C}$, $\mathbf{X}_{I}$}
\nonl Offline Learning\\
Calculate $\mathcal{X}_I$ by \eqref{eq:P-GC-C-obj}-\eqref{eq:P-GC-C-con3}\;
Calculate $\{\mathbf{c}_1^\star,\ldots,\mathbf{c}_K^\star\}$ by iteratively solving \eqref{eq:BV}, with $K=2^{B_V}$\;
$\mathcal{C}\leftarrow\{\mathbf{c}_{1}^\star,\mathbf{c}_{2}^\star,\ldots,\mathbf{c}_{K}^\star\}$\;
\nonl Online Deployment\\
Initialize $\mathbf{X}_{I,g}=\mathbf{0},\:\forall g$\;
\While{no convergence of objective \eqref{eq:P-GC-D-obj4}}{
\For{$g\leftarrow 1$ \KwTo $G$}{
Update $\mathbf{X}_{I,g}\in \mathcal{C}_V$ by searching among the $2^{B_V}$ possible values\;
}
}
\caption{Vector-Discrete RISs Design}
\label{alg:vector}
\end{algorithm}

The cost of vector-discrete RISs relies on their optimization complexity.
For example, if $B$ bits are allocated to each reactance element, the number of clusters needed to quantize a vector of $N_{G}\left(N_{G}+1\right)/2$ elements is $k=2^{B\frac{N_{G}\left(N_{G}+1\right)}{2}}$, growing exponentially with the square of $N_{G}$.
Consequently, the number of search times in a complete iteration of the alternating optimization algorithm becomes $G2^{B\frac{N_{G}\left(N_{G}+1\right)}{2}}$.
However, this discretization strategy based on vector quantization brings two benefits.
First, for the same number of total resolution bits, vector quantization is inherently more efficient than scalar quantization.
Second, the number of total resolution bits can be chosen with more degrees of freedom, since it is no longer limited to $B$ bits for each reactance element.

\section{Performance Evaluation}
\label{sec:results}

In this section, we evaluate the performance of an RIS-aided \gls{mimo} system with different discrete RIS configurations, each characterized by its number of elements $N_{I}$, number of resolution bits, and group size $N_{G}$.
The performance is measured in terms of received signal power, given by $P_{R}=P_{T}\left\|\mathbf{H}_{RT}+\mathbf{H}_{RI}\boldsymbol{\Theta}\mathbf{H}_{IT}\right\|^{2}$.

\begin{figure*}[t]
    \begin{centering} 
    \includegraphics[width=0.24\textwidth]{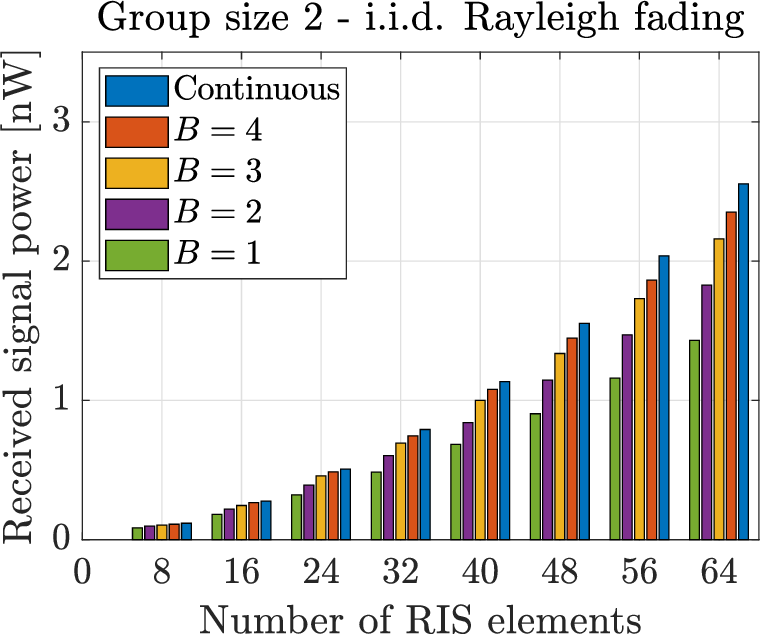}
    \includegraphics[width=0.24\textwidth]{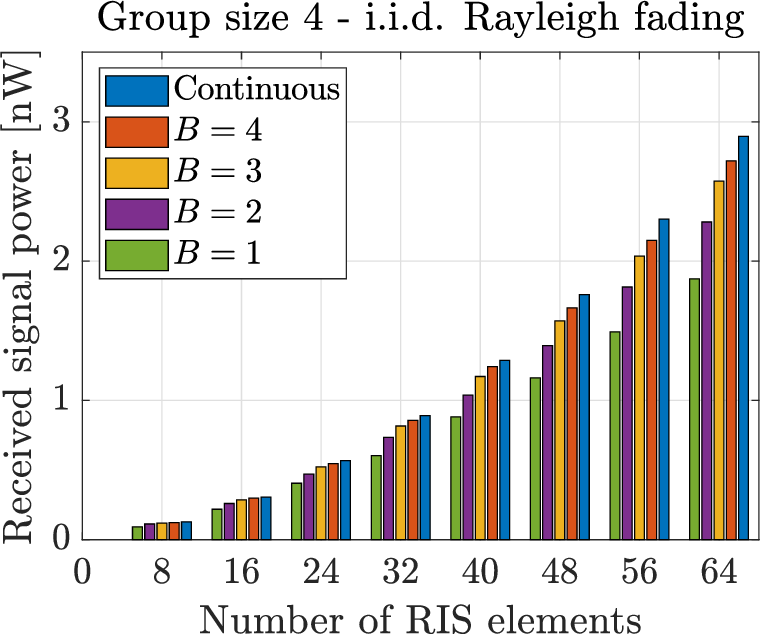}
    \includegraphics[width=0.24\textwidth]{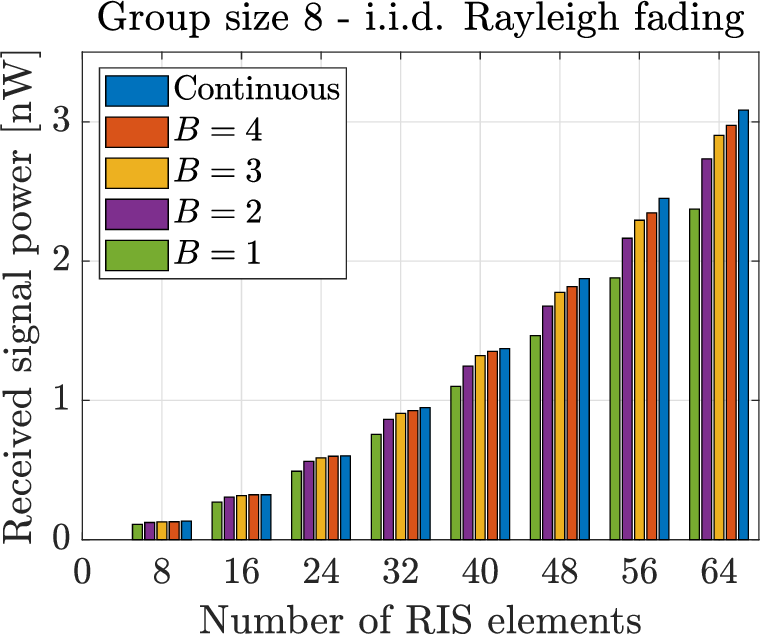}
    \includegraphics[width=0.24\textwidth]{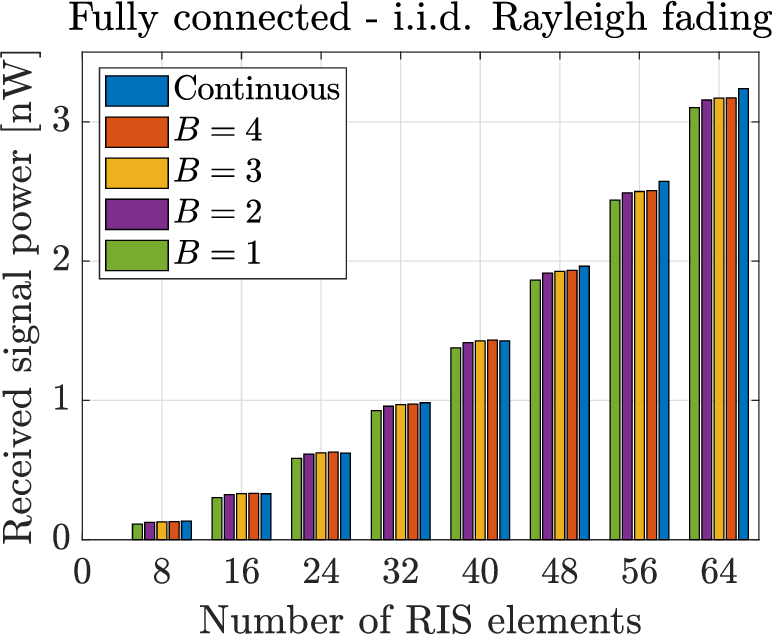}
    \par\end{centering}
    \vspace{0.1cm}
    \begin{centering}
    \includegraphics[width=0.24\textwidth]{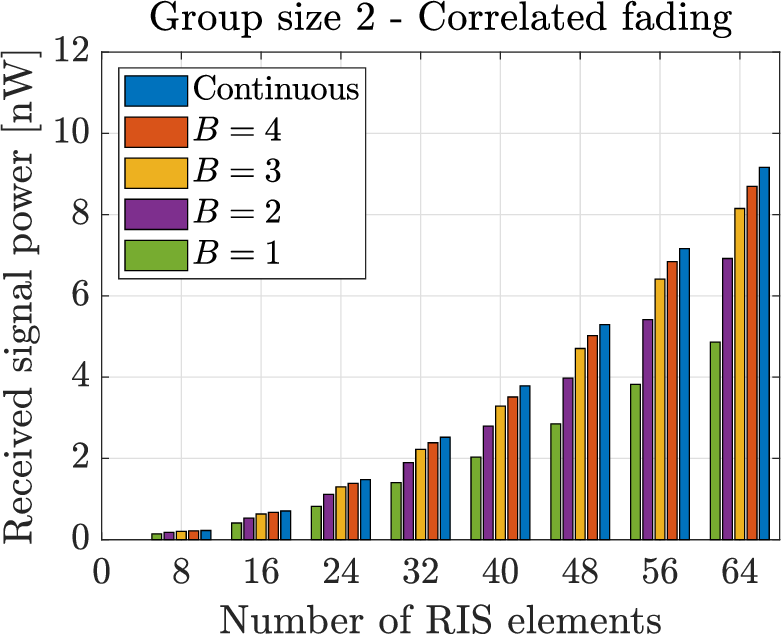}
    \includegraphics[width=0.24\textwidth]{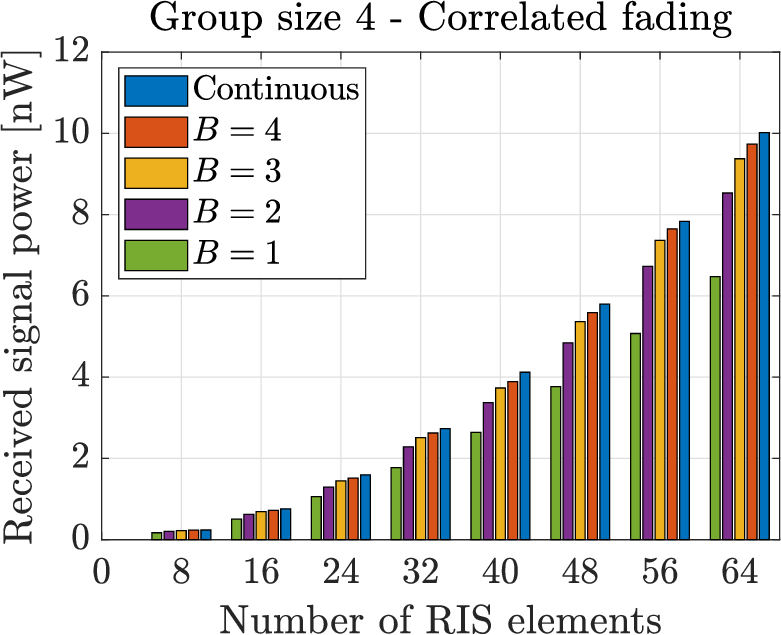}
    \includegraphics[width=0.24\textwidth]{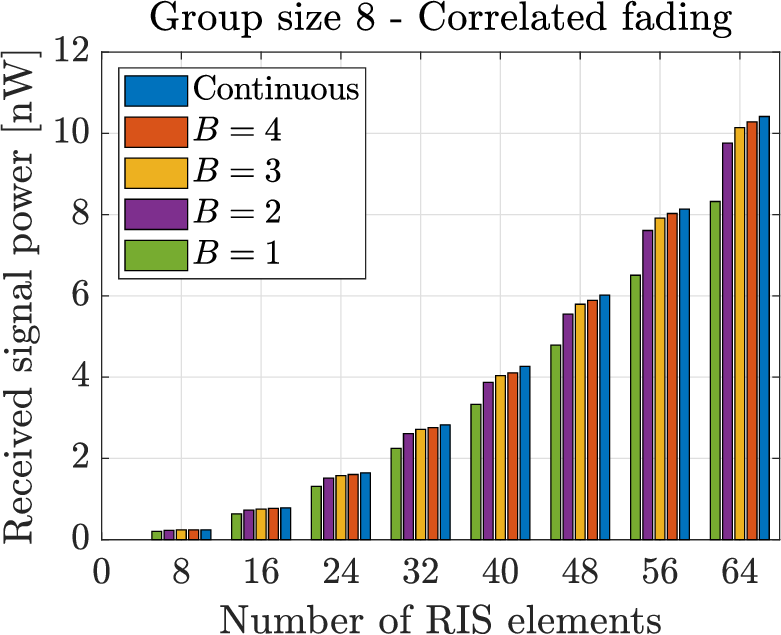}
    \includegraphics[width=0.24\textwidth]{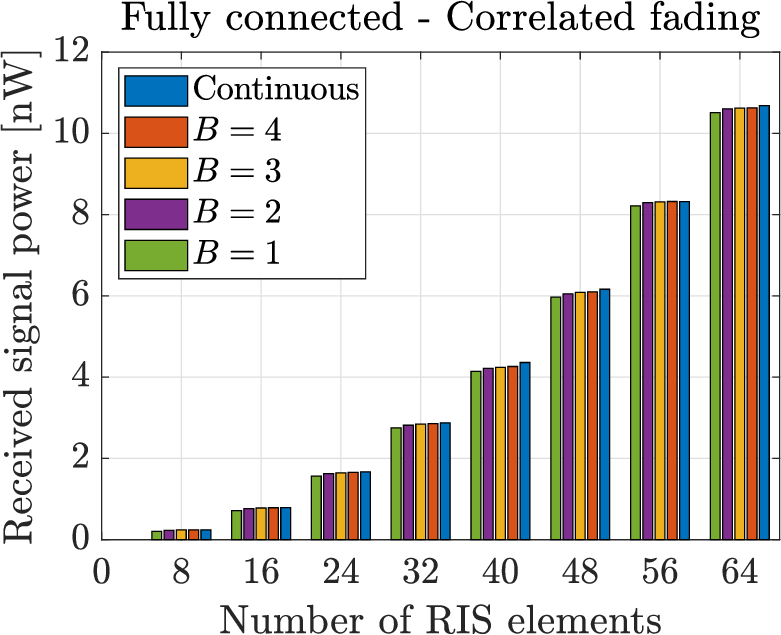}
    \par\end{centering}
    \caption{Average received signal power versus the number of RIS elements for different values of group size.}
    \label{fig:group-size}
\end{figure*}

\subsection{Numerical Simulation Setup}

Let us consider a three-dimensional coordinate system $(x,y,z)$, in which the $z$-axis represents the height above the ground in meters (m).
The transmitter, the RIS, and the receiver are \glspl{ula} composed of $N_{T}$, $N_{I}$, and $N_{R}$ antennas, respectively, with half wavelength antenna spacing\footnote{Our discrete design for BD-RIS is valid for any RIS array shape.}.
We set $N_{T}=4$ and $N_{R}=2$, while considering several values of $N_{I}$ as specified in the following.
The transmitter is located at $(5,-250,25)$ and its antennas are arranged along the $x$-axis.
The RIS is located at $(0,0,5)$ and its antennas are arranged along the $y$-axis.
Finally, the receiver is located at $(5,5,1.5)$ with random orientation.
The path loss is modeled as $L_{ij}(d_{ij})=L_{0}d_{ij}^{-\alpha_{ij}}$, where $L_{0}$ is the path loss at distance 1 m, $d_{ij}$ is the distance, and $\alpha_{ij}$ is the path loss exponent for $ij\in\{RT,RI,IT\}$.
We set $L_{0}=-30$ dB, $\alpha_{RT}=4$, $\alpha_{RI}=2.8$, $\alpha_{IT}=2$, and $P_{T}=10$ W. 

We analyze two scenarios with different small-scale fading effects: i.i.d. Rayleigh fading and correlated fading.
In the former scenario, the small-scale fading of all channels is assumed to be i.i.d. Rayleigh distributed.
Thus, we have $\mathbf{H}_{RT}\sim\mathcal{CN}\left(\boldsymbol{0},L_{RT}\mathbf{I}\right)$, $\mathbf{H}_{RI}\sim\mathcal{CN}\left(\boldsymbol{0},L_{RI}\mathbf{I}\right)$ and $\mathbf{H}_{IT}\sim\mathcal{CN}\left(\boldsymbol{0},L_{IT}\mathbf{I}\right)$.
In the latter scenario, we generate the small-scale effects with QuaDRiGa version 2.4, a \textsc{Matlab} based statistical ray-tracing channel simulator \cite{jae14}.
An urban macrocell propagation environment is simulated, with \gls{los} in the transmitter-RIS link, and with \gls{nlos} in the transmitter-receiver and RIS-receiver links.
The channel models ``3GPP\_38.901\_UMa\_LOS'' and ``3GPP\_38.901\_UMa\_NLOS'' have been used to simulate the \gls{los} and \gls{nlos} channels, respectively.
The \glspl{ula} at the transmitter, the RIS, and the receiver are modeled as ``3gpp-3d'' array and the carrier frequency is set to $f_c=2.5$ GHz.
%

For both scenarios, offline learning stages have been carried out to define the optimal codebooks.
We employ a training set $\mathcal{H}_0$ composed of $N_0=100$ channel realization triplets for scalar-discrete RISs, and $N_0=500$ triplets for vector-discrete RISs.
These values of $N_0$ have been set such that in the $K$-means clustering problems involved in Alg.~\ref{alg:scalar} and Alg.~\ref{alg:vector}, the number of training samples is always higher than the number of clusters.
In the alternating optimizations, convergence is considered reached when the fractional increase of the objective value is below $\epsilon=10^{-3}$.
Finally, the average received signal power has been computed for each RIS configuration using the Monte Carlo method.

\subsection{Comparison With Single Connected RIS}

Before comparing discrete-value group and fully connected RISs with discrete-value single connected RISs, we briefly review how single connected RISs have been discretized in the literature.
In single connected RISs, $\boldsymbol{\Theta}$ is completely described by the phase shifts $\theta_{n_{I}}$.
The phase shifts have been typically discretized uniformly in $\left[0,2\pi\right)$.
By assigning $B$ resolution bits per phase shift, they are chosen in the codebook
\begin{equation}
\mathcal{C}^{\text{Single}} = \left\{ 0,\delta,\ldots,\left( 2^{B} - 1 \right) \delta \right\},
\end{equation}
where $\delta = \frac{2\pi}{2^{B}}$ is the quantization step.
Thus, the received signal power maximization problem is formulated in the discrete case as
\begin{align}
\underset{\theta_{n_{I}}}{\mathsf{\mathrm{max}}}\;\;
& \left\|\mathbf{H}_{RT}+\mathbf{H}_{RI}\boldsymbol{\Theta}\mathbf{H}_{IT}\right\|^{2}\label{eq:P-SC-D-obj}\\
\mathsf{\mathrm{s.t.}}\;\;\;
& \boldsymbol{\Theta}=\mathrm{diag}\left(e^{j\theta_{1}},\ldots,e^{j\theta_{N_{I}}}\right),\label{eq:P-SC-D-con1}\\
& \theta_{n_{I}}\in \left\{ 0,\delta,\ldots,\left( 2^{B} - 1 \right) \delta \right\},\:\forall n_{I}.\label{eq:P-SC-D-con2}
\end{align}
To solve this problem, we employ the widely adopted alternating optimization method \cite{you20}, \cite{abe20}, \cite{wu19b}.
In this method, each of the $N_I$ phase shifts is optimized individually by searching among all the possible $2^B$ values, while fixing the other $N_I-1$ phase shifts.
This procedure is repeated for all $N_I$ phase shifts, iterating multiple times until the convergence is reached, i.e., until the fractional increase of the objective value is below a certain parameter $\epsilon$.

\begin{figure*}[t]
    \centering
    \includegraphics[height=0.3\textwidth]{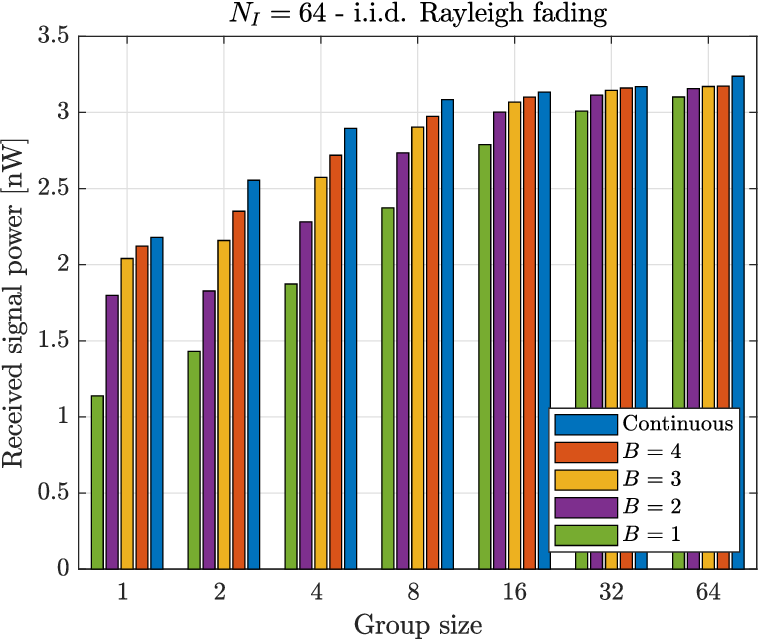}
    \includegraphics[height=0.3\textwidth]{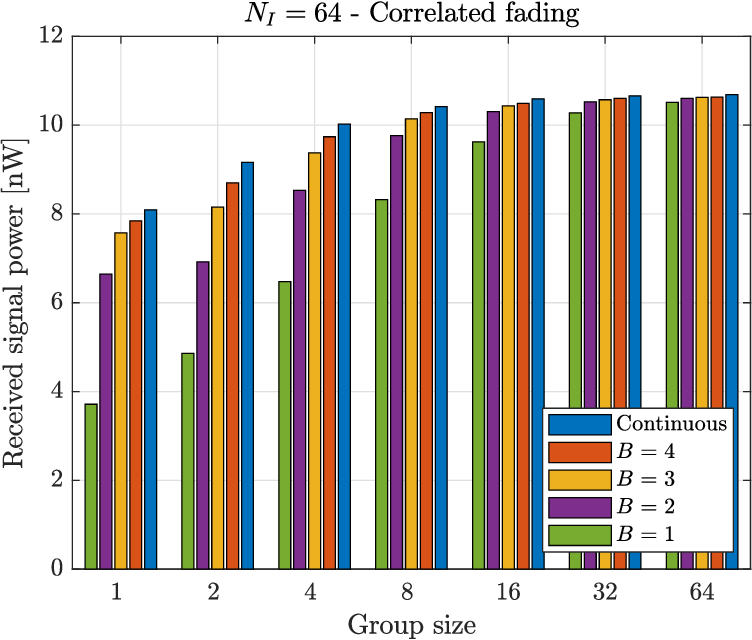}
    \caption{Average received signal power versus the group size.}
    \label{fig:ni64}
\end{figure*}
\begin{figure*}[t]
    \centering
    \includegraphics[height=0.3\textwidth]{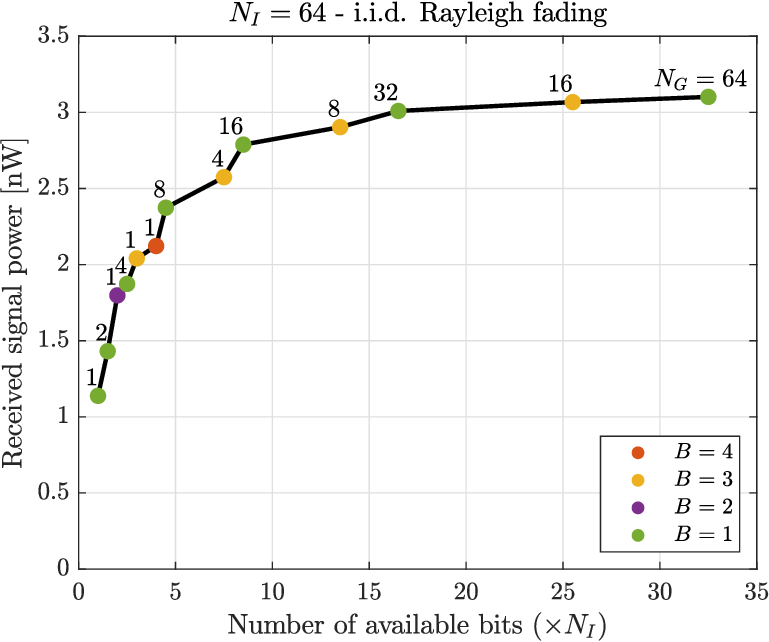}
    \includegraphics[height=0.3\textwidth]{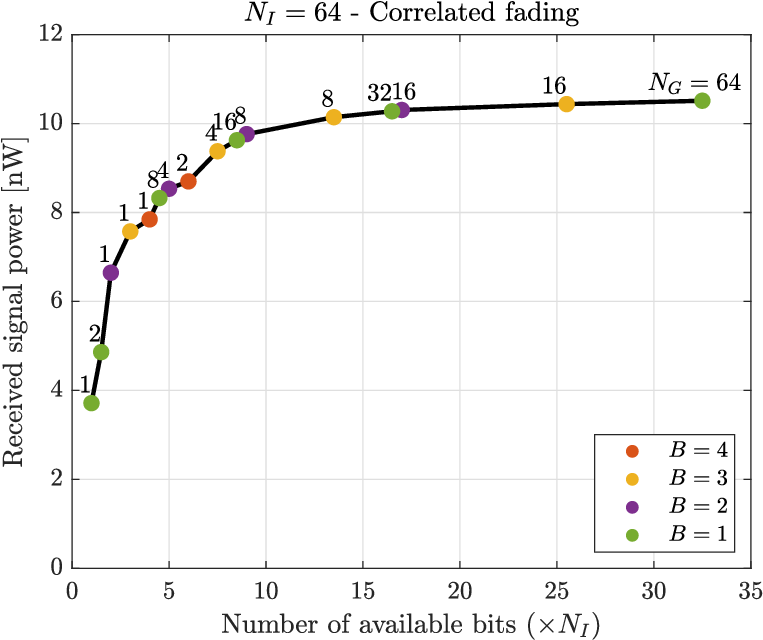}
    \caption{Average received signal power versus the number of total resolution bits.}
    \label{fig:ni64-vs-bits}
\end{figure*}

\subsection{Scalar-Discrete Group/Fully Connected RIS}

We first evaluate the received signal power in the case of scalar-discrete RISs.
In Fig.~\ref{fig:bit}, we show  the received signal power for different values of $N_{I}$, using from one to four resolution bits per reactance element.
Since this is the first work investigating the design of BD-RISs based on discrete values, the only available benchmark is given by the performance achieved by the same BD-RISs optimized with continuous values.
In each subfigure, we compare single connected RISs, group connected RIS with $N_{G}\in\{2,4,8\}$, and fully connected RISs.
The fully connected architecture achieves the highest received signal power, as in the case of continuous-value RISs \cite{she20}.
We also observe that a larger group size yields a higher received signal power.
Thus, group connected RISs always perform better than the conventional single connected RISs.
Furthermore, the discrepancy in the performance of fully and single connected is higher when fewer resolution bits are employed.
With correlated small-scale fading, the received signal power is higher than with i.i.d. Rayleigh fading channels.
This is because dominant eigenmode transmission achieves a received signal power proportional to the dominant eigenvalue of $\mathbf{H}\mathbf{H}^H$.
Thus, correlated channels, being low-rank, are beneficial compared to independent channels.

In Fig.~\ref{fig:group-size}, we show the received signal power when group connected (with group size $N_{G}\in\{2,4,8\}$) and fully connected RISs are used in the presence of i.i.d. Rayleigh fading channels and correlated channels.
The performance of scalar-discrete RISs is compared with the performance of continuous-value RISs, optimized by solving \eqref{eq:P-GC-C-obj}-\eqref{eq:P-GC-C-con3}.
We observe that the higher the group size, the fewer resolution bits are necessary to reach the performance of continuous-value RISs.
The rationale behind this behavior is given by the necessary and sufficient condition for $\boldsymbol{\Theta}$ to achieve the maximum $P_R$ with no direct link.
By extending the analysis in \cite{she20} to \gls{mimo} settings, it is possible to prove that this condition is given by
\begin{equation}
\mathbf{u}_{RI,g}=\boldsymbol{\Theta}_{g}\mathbf{u}_{IT,g},\forall g,
\label{eq:opt-cond-u}
\end{equation}
where $\mathbf{u}_{RI}=\left[\mathbf{u}_{RI,1},\mathbf{u}_{RI,2},\ldots,\mathbf{u}_{RI,G}\right]$ with $\mathbf{u}_{RI,g}\in\mathbb{C}^{N_{G}\times1}$, and $\mathbf{u}_{IT}=\left[\mathbf{u}_{IT,1},\mathbf{u}_{IT,2},\ldots,\mathbf{u}_{IT,G}\right]^{T}$ with $\mathbf{u}_{IT,g}\in\mathbb{C}^{N_{G}\times1}$, are the dominant left singular vectors of $\mathbf{H}_{RI}^{H}$ and $\mathbf{H}_{IT}$, respectively\footnote{
Condition \eqref{eq:opt-cond-u} is derived by considering $P_{R}\leq P_{T}\underset{\left\|\mathbf{x}\right\|=1}{\mathsf{\mathrm{max}}}\;\left\|\mathbf{H}_{RI}\right\|^{2}\left\|\boldsymbol{\Theta}\mathbf{H}_{IT}\mathbf{x}\right\|^{2}\leq P_{T}\underset{\left\|\mathbf{x}\right\|=1}{\mathsf{\mathrm{max}}}\;\left\|\mathbf{H}_{RI}\right\|^{2}\left\|\boldsymbol{\Theta}\mathbf{H}_{IT}\right\|^{2}\left\|\mathbf{x}\right\|^{2}$.
Note that the equality holds in the two inequalities when $\mathbf{u}_{RI}$ is equal to the dominant left singular vector of $\boldsymbol{\Theta}\mathbf{H}_{IT}$, i.e., $\boldsymbol{\Theta}\mathbf{u}_{IT}$, which is equivalent to \eqref{eq:opt-cond-u} in the case of block diagonal $\boldsymbol{\Theta}$.}.
This condition consists of an underdetermined system of $N_{I}$ equations in $N_{I}\left(N_{G}+1\right)/2$ unknowns.
Thus, a higher group size $N_{G}$ implies more degrees of freedom, and fewer resolution bits are required to satisfy it.
In particular, in the fully connected architecture, a single resolution bit is sufficient to obtain approximately the same performance as the optimized continuous-value RISs.
%
It is worthwhile to clarify the difference between this conclusion and the results in \cite{jun21}.
In \cite{jun21}, the authors prove that single connected RISs with discrete phase shifts can achieve an \gls{snr} in the order of $\mathcal{O}\left(N_{I}^2\right)$, regardless the number of bits $B\geq 1$ (see Proposition 1 in \cite{jun21}).
This means that a single resolution bit ($B = 1$) can achieve the same performance growth as $B=\infty$ in the limit $N_{I}\rightarrow\infty$.
However, when a practical number of RIS elements is considered, $B = 1$ cannot achieve the same performance as $B=\infty$.
Furthermore, even if $B = 1$ and $B=\infty$ achieve the same asymptotic \gls{snr} growth, $B = 1$ causes a power loss of 3.9 dB \cite{wu19b}.
Conversely, in this work, we show that $B = 1$ can achieve approximately the same performance as $B=\infty$ in fully connected architectures for any $N_{I}$, as represented in Fig.~\ref{fig:group-size}.

\begin{figure*}[t]
    \centering
    \includegraphics[height=0.3\textwidth]{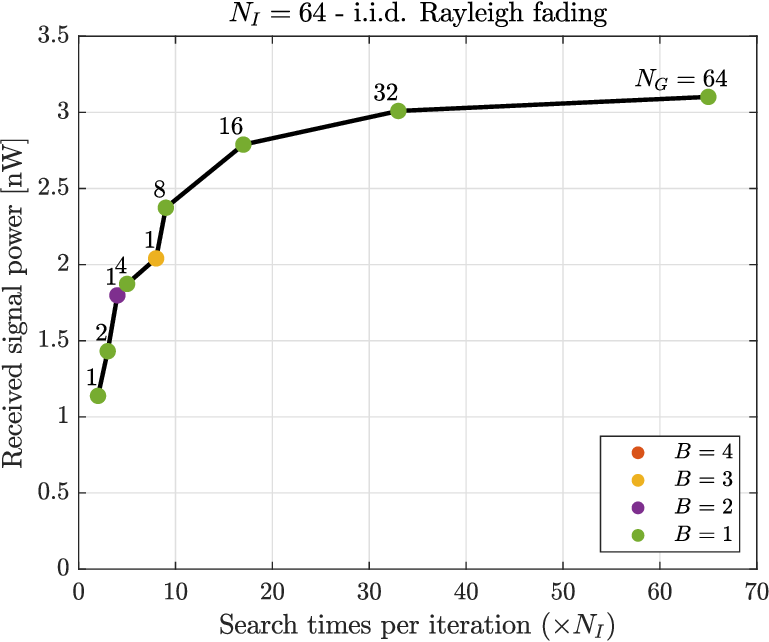}
    \includegraphics[height=0.3\textwidth]{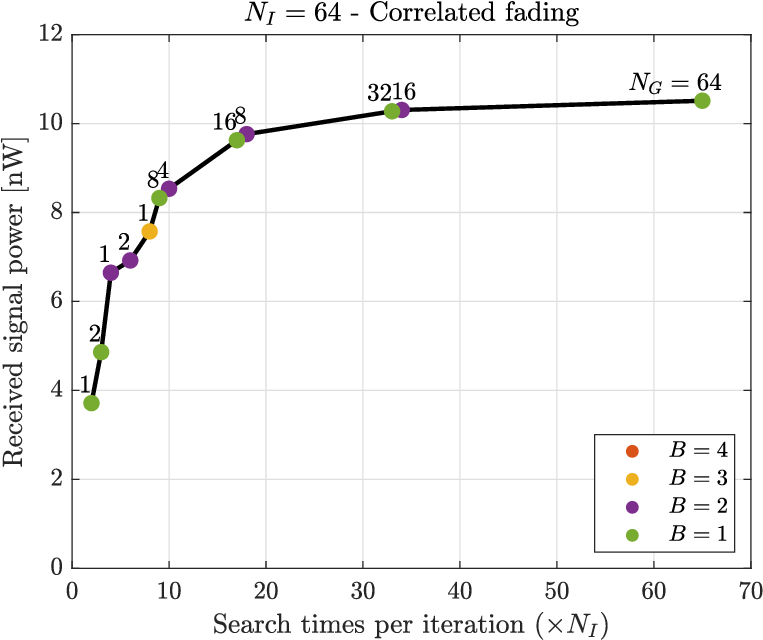}
    \caption{Average received signal power versus the computational complexity.}
    \label{fig:ni64-vs-compl}
\end{figure*}

We now compare the performance obtained with different group sizes when the number of RIS elements $N_{I}$ is fixed.
To this end, we optimize RISs with $N_{I}=64$ elements considering all the possible group sizes, spanning from the single connected architecture ($N_{G}=1$) to the fully connected ($N_{G}=64$).
In Fig.~\ref{fig:ni64}, the upper bound of the received signal power is reported together with the power maximized with different resolutions.
We notice that the received signal power increases with the group size for every $B$ considered.
Furthermore, the performance of discrete-value RISs with $B=1$ approaches the performance of continuous-value RISs as the group size increases, converging to it in the fully connected architecture.

When $B$ bits are allocated to each $\mathbf{X}_{I}$ element, the number of total bits employed scales with the number of $\mathbf{X}_{I}$ elements as $B\frac{\left(N_{G}+1\right)}{2}N_{I}$.
Fixing $N_{I}=64$, we now investigate which are the RIS configurations, described by the pairs $(N_{G},B)$, that maximize the received signal power when the number of total bits is limited.
Fig.~\ref{fig:ni64-vs-bits} shows how the single, group, and fully connected architectures with discrete reconfigurable impedances can provide the compromise between performance and hardware complexity.
Here, each point represents the received signal power achievable with a specific RIS configuration, when $B\frac{\left(N_{G}+1\right)}{2}\times 64$ total resolution bits are available.
Furthermore, the numeric labels above the points identify the group size $N_{G}$.
We observe that RIS configurations with small values of $B$ are selected as optimal in case of a limited number of total bits.
Therefore, RISs with larger group sizes and lower resolution are generally preferred over RISs with smaller group sizes and higher resolution.
This result confirms the effectiveness of group and fully connected RISs.

We now discuss the trade-off between computational complexity and performance.
To this end, we analyze the optimization computational complexity and the received signal power of all RIS configurations with $N_I=64$ elements, each described by the group size $N_G$ and the number of resolution bits $B$.
In Fig.~\ref{fig:ni64-vs-compl}, we report the RIS architectures achieving the most favorable trade-off between performance and computational complexity. 
Here, each point represents the received signal power achievable with a specific RIS configuration, requiring $N_{I}\frac{\left(N_{G}+1\right)}{2}2^{B}$ search times per iteration of our optimization algorithm.
Furthermore, the numeric labels above the points identify the group size $N_G$.
We observe that RIS configurations with small values of $B$ are the ones offering the best trade-off between performance and computational complexity.
This result confirms the effectiveness of group and fully connected RISs over single connected RISs also under the constraint of discrete values.

\begin{figure*}[t]
    \centering
    \includegraphics[height=0.3\textwidth]{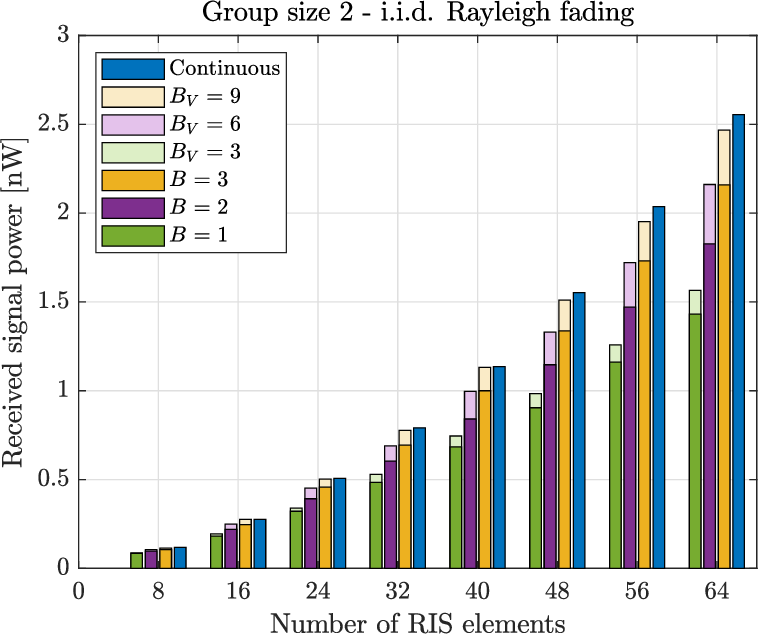}
    \includegraphics[height=0.3\textwidth]{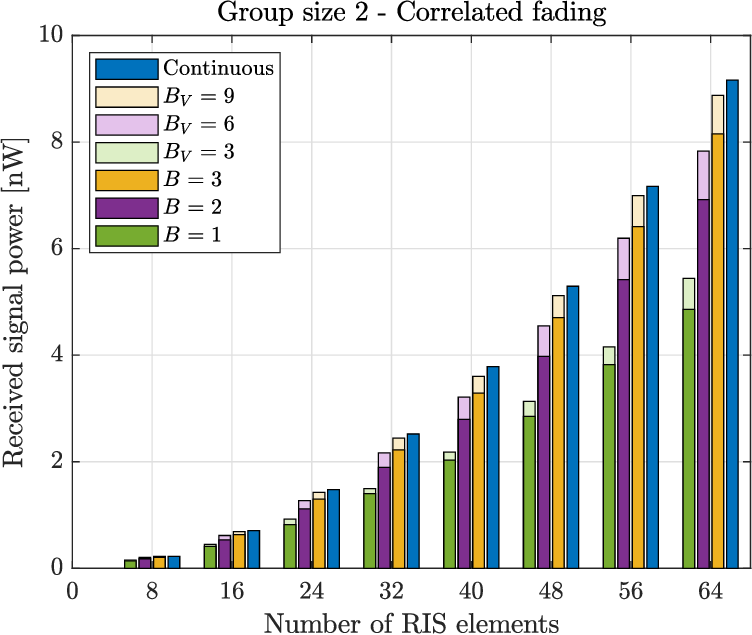}
    \caption{Average received signal power versus the number of RIS elements in the scalar-discrete and vector-discrete group connected architectures.}
    \label{fig:vector-discrete-ng2}
\end{figure*}
\begin{figure*}[t]
    \centering
    \includegraphics[height=0.3\textwidth]{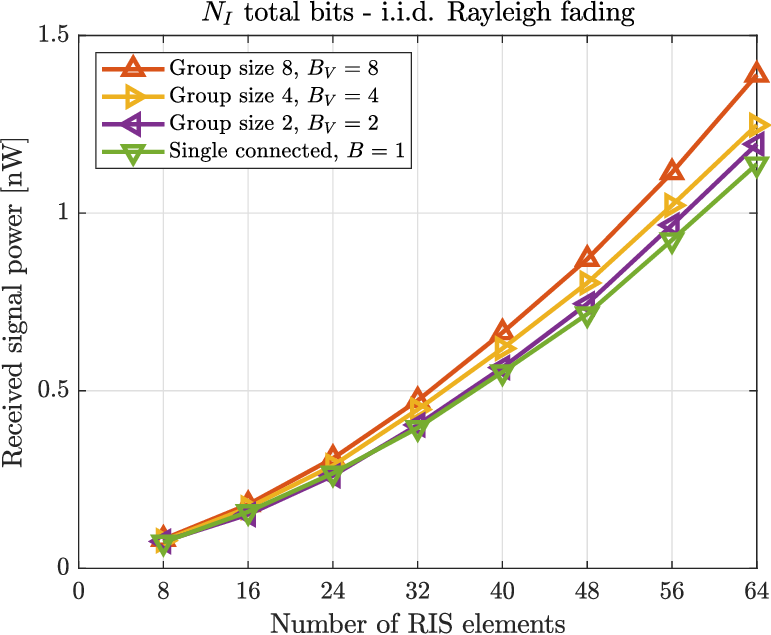}
    \includegraphics[height=0.3\textwidth]{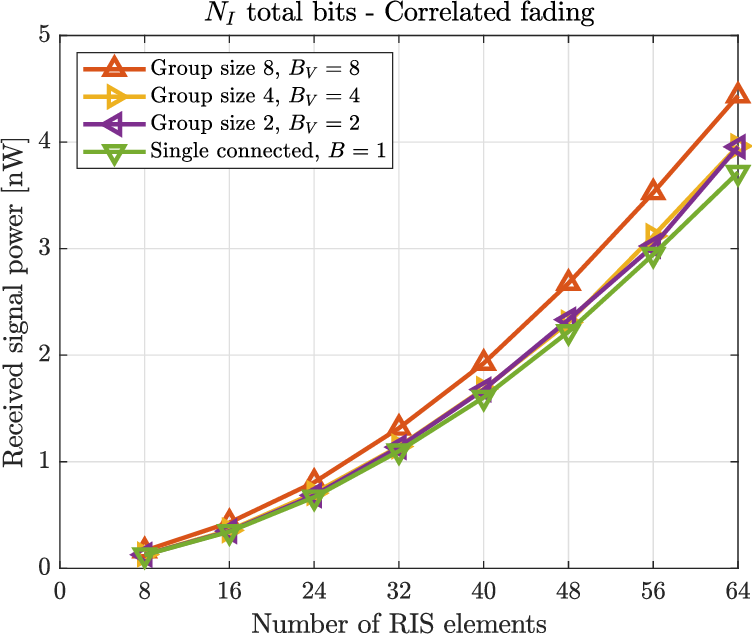}
    \caption{Average received signal power versus the number of RIS elements for different group sizes, with $N_{I}$ total resolution bits.}
    \label{fig:vector-discrete-B0}
\end{figure*}

\subsection{Vector-Discrete Group/Fully Connected RIS}

Let us analyze the performance of vector-discrete RISs, in comparison with scalar-discrete RISs.
To this end, we set the number of resolution bits assigned to each block $\mathbf{X}_{I,g}$ to $B_V=B\frac{N_{G}\left(N_{G}+1\right)}{2}$.
This is equivalent, in terms of total bits employed, to assigning $B$ bits to each reactance element, allowing a comparison between scalar-discrete and vector-discrete RISs.
Fig.~\ref{fig:vector-discrete-ng2} shows the performance achieved by vector-discrete group connected RISs with group size $N_{G}=2$, when the number of resolution bits considered is $B_V\in\{3,6,9\}$.
In terms of total resolution bits employed, $B_V\in\{3,6,9\}$ bits per impedance network block are equivalent to $B\in\{1,2,3\}$ bits per impedance element, respectively.
Thus, for the sake of comparison, we report in Fig.~\ref{fig:vector-discrete-ng2} also the performance of scalar-discrete group connected RISs with $B\in\{1,2,3\}$.
We notice that vector discretization brings a significant performance improvement over scalar discretization.
In particular, $B_V=9$ resolution bits per impedance network block are sufficient to approach the performance of continuous-value RISs.
However, $B=3$ resolution bits per impedance element are not sufficient to reach the performance of continuous-value RISs when scalar discretization is considered.

We now compare the performance of RISs with different group sizes when the same number of total bits is employed.
To this end, we fix the number of total bits used to $N_{I}$, which are required by the single connected architecture with $B=1$.
To use $N_{I}$ total bits, the feature space having $N_{G}\left(N_{G}+1\right)/2$ dimensions is quantized with $B_{V}=N_{G}$ resolution bits.
Fig.~\ref{fig:vector-discrete-B0} shows the received signal power achieved with different values of group size.
Here, we can notice that higher group sizes achieve better performance than lower group sizes.
This is due to the optimality condition \eqref{eq:opt-cond-u}, which has more degrees of freedom as the group size increases.
Thus, it can be more easily satisfied when higher group sizes are considered.

\begin{figure*}[t]
    \centering
    \includegraphics[height=0.3\textwidth]{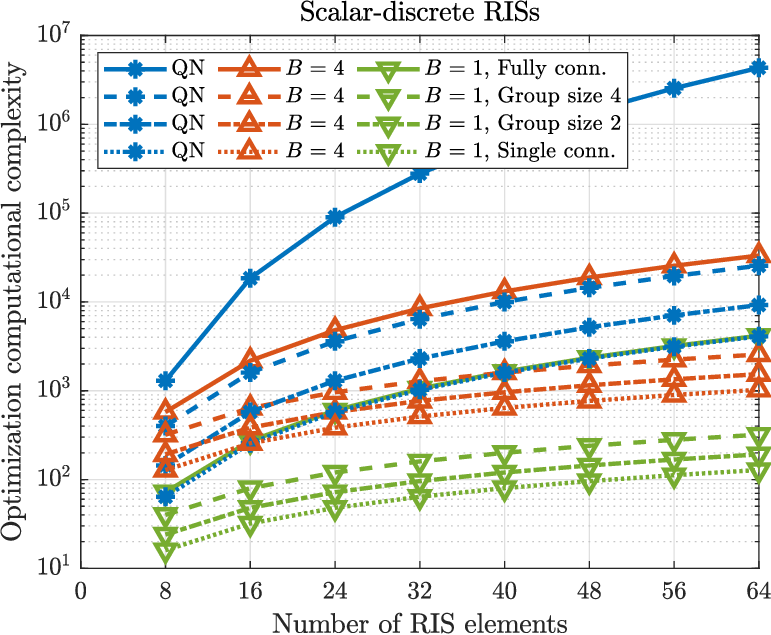}
    \includegraphics[height=0.3\textwidth]{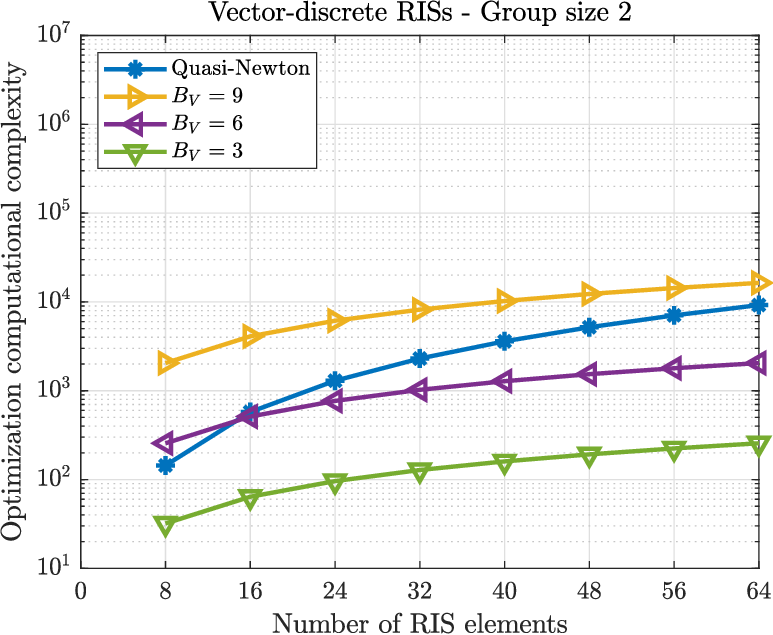}
    \caption{Optimization computational complexity versus the number of RIS elements. The quasi-Newton method ``QN'' is compared with alternating optimization for scalar-discrete RISs (on the left) and vector-discrete RISs (on the right).}
    \label{fig:complexity}
\end{figure*}

\subsection{Optimization Computational Complexity}

To conclude our discussion, we compare continuous-value RISs with discrete-value RISs in terms of optimization computational complexity.
On the one hand, when the quasi-Newton method is used to optimize the continuous values of $\mathbf{X}_{I}$, the computational complexity of each iteration is $\mathcal{O}((N_{I}(N_{G}+1)/2)^2)$ \cite{she20}.
On the other hand, when alternating optimization is used to optimize the discrete values of $\mathbf{X}_{I}$, we distinguish two cases, namely scalar-discrete RISs and vector-discrete RISs.
In the case of scalar-discrete RISs, the total number of search times in a complete iteration of the alternating optimization algorithm is $G\frac{N_{G}\left(N_{G}+1\right)}{2}2^{B}$ in group connected architectures.
This number boils down to $N_I2^B$ in single connected and to $\frac{N_{I}\left(N_{I}+1\right)}{2}2^{B}$ in fully connected architectures.
Instead, in the case of vector-discrete RISs, the number of search times per iteration becomes $\frac{N_{I}}{N_{G}}2^{B_V}$.
The values of these computational complexities are reported in Fig.~\ref{fig:complexity}.
In the case of scalar-discrete RISs, we observe that our RIS design strategy based on discrete-value impedance networks is beneficial also in terms of optimization computational complexity.
Even with $B=4$, the complexity is far less than the complexity of the quasi-Newton method for medium-high numbers of RIS elements.

Remarkably, the complexity of our algorithm grows with $\mathcal{O}(N_I)$ in the case of scalar-discrete group connected RISs, and with $\mathcal{O}(N_I^2)$ in the case of scalar-discrete fully connected RISs.
Besides, the complexity grows with $\mathcal{O}(N_I)$ in the case of vector-discrete group connected RISs, and does not depend on $N_I$ in the case of vector-discrete fully connected RISs.
Thus, we can conclude that our algorithm is scalable and can be applied to RIS with a high number of elements $N_I$.

\section{Conclusion}
\label{sec:conclusion}

We propose a novel BD-RIS design based on discrete-value group and fully connected architectures.
Our strategy is composed of two stages.
Firstly, through the offline learning stage, we build the codebook containing the possible values for the reconfigurable impedances.
Secondly, during the online deployment stage, we optimize the reconfigurable impedances with alternating optimization on the basis of the designed codebook.
Two different approaches are developed, namely scalar-discrete and vector-discrete RISs, based on scalar and vector quantization, respectively.
Vector-discrete RISs achieve higher performance than scalar-discrete RISs at the cost of increased optimization computational complexity.
Numerical results show that a single resolution bit per reconfigurable impedance is sufficient to achieve optimality in fully connected RISs.
In a practical scenario, this simplifies significantly the hardware complexity of the fully connected RIS.

Two future research directions can be identified.
First, the optimization of discrete-value BD-RIS based on imperfect or statistical channel knowledge should be investigated.
Second, the proposed codebook design and optimization framework can be readily extended to multi-user systems.
However, the involved optimization problems are non-convex and hard to solve.
Their solution could be investigated in future work.

\bibliographystyle{IEEEtran}
\bibliography{IEEEabrv,main}

\end{document}

%% file: glossary.tex
\newacronym{ao}{AO}{alternating optimization}
\newacronym{mimo}{MIMO}{multiple-input multiple-output}
\newacronym{siso}{SISO}{single-input single-output}
\newacronym{awgn}{AWGN}{additive white Gaussian noise}
\newacronym{csi}{CSI}{channel state information}
\newacronym{ula}{ULA}{uniform linear array}
\newacronym{upa}{UPA}{uniform planar array}
\newacronym{swipt}{SWIPT}{simultaneous wireless information and power transfer}
\newacronym{oma}{OMA}{orthogonal multiple access}
\newacronym{noma}{NOMA}{non-orthogonal multiple access}
\newacronym{sinr}{SINR}{signal-to-interference-plus-noise ratio}
\newacronym{snr}{SNR}{signal-to-noise ratio}
\newacronym{los}{LoS}{line-of-sight}
\newacronym{nlos}{NLoS}{non-line-of-sight}
\newacronym{rf}{RF}{radio frequency}
\newacronym{mad}{MAD}{median absolute deviation}
\newacronym{rsma}{RSMA}{rate splitting multiple access}
\newacronym{star-ris}{STAR-RIS}{simultaneously transmitting and reflecting RIS}

%% file: main.bbl
\begin{thebibliography}{10}
\providecommand{\url}[1]{#1}
\csname url@samestyle\endcsname
\providecommand{\newblock}{\relax}
\providecommand{\bibinfo}[2]{#2}
\providecommand{\BIBentrySTDinterwordspacing}{\spaceskip=0pt\relax}
\providecommand{\BIBentryALTinterwordstretchfactor}{4}
\providecommand{\BIBentryALTinterwordspacing}{\spaceskip=\fontdimen2\font plus
\BIBentryALTinterwordstretchfactor\fontdimen3\font minus
  \fontdimen4\font\relax}
\providecommand{\BIBforeignlanguage}[2]{{%
\expandafter\ifx\csname l@#1\endcsname\relax
\typeout{** WARNING: IEEEtran.bst: No hyphenation pattern has been}%
\typeout{** loaded for the language `#1'. Using the pattern for}%
\typeout{** the default language instead.}%
\else
\language=\csname l@#1\endcsname
\fi
#2}}
\providecommand{\BIBdecl}{\relax}
\BIBdecl

\bibitem{bas19}
E.~Basar, M.~Di~Renzo, J.~De~Rosny, M.~Debbah, M.-S. Alouini, and R.~Zhang,
  ``Wireless communications through reconfigurable intelligent surfaces,''
  \emph{IEEE Access}, vol.~7, pp. 116\,753--116\,773, 2019.

\bibitem{wu19a}
Q.~Wu and R.~Zhang, ``Towards smart and reconfigurable environment: Intelligent
  reflecting surface aided wireless network,'' \emph{IEEE Communications
  Magazine}, vol.~58, no.~1, pp. 106--112, 2020.

\bibitem{liu21}
Y.~Liu, X.~Liu, X.~Mu, T.~Hou, J.~Xu, M.~Di~Renzo, and N.~Al-Dhahir,
  ``Reconfigurable intelligent surfaces: Principles and opportunities,''
  \emph{IEEE Communications Surveys Tutorials}, vol.~23, no.~3, pp. 1546--1577,
  2021.

\bibitem{wu21}
Q.~Wu, S.~Zhang, B.~Zheng, C.~You, and R.~Zhang, ``Intelligent reflecting
  surface-aided wireless communications: A tutorial,'' \emph{IEEE Transactions
  on Communications}, vol.~69, no.~5, pp. 3313--3351, 2021.

\bibitem{bad19}
M.-A. Badiu and J.~P. Coon, ``Communication through a large reflecting surface
  with phase errors,'' \emph{IEEE Wireless Communications Letters}, vol.~9,
  no.~2, pp. 184--188, 2020.

\bibitem{xu21}
P.~Xu, G.~Chen, Z.~Yang, and M.~D. Renzo, ``Reconfigurable intelligent
  surfaces-assisted communications with discrete phase shifts: How many
  quantization levels are required to achieve full diversity?'' \emph{IEEE
  Wireless Communications Letters}, vol.~10, no.~2, pp. 358--362, 2021.

\bibitem{zha20b}
H.~Zhang, B.~Di, L.~Song, and Z.~Han, ``Reconfigurable intelligent surfaces
  assisted communications with limited phase shifts: How many phase shifts are
  enough?'' \emph{IEEE Transactions on Vehicular Technology}, vol.~69, no.~4,
  pp. 4498--4502, 2020.

\bibitem{li20}
D.~Li, ``Ergodic capacity of intelligent reflecting surface-assisted
  communication systems with phase errors,'' \emph{IEEE Communications
  Letters}, vol.~24, no.~8, pp. 1646--1650, 2020.

\bibitem{liu20}
S.~Liu, Z.~Gao, J.~Zhang, M.~D. Renzo, and M.-S. Alouini, ``Deep denoising
  neural network assisted compressive channel estimation for mmwave intelligent
  reflecting surfaces,'' \emph{IEEE Transactions on Vehicular Technology},
  vol.~69, no.~8, pp. 9223--9228, 2020.

\bibitem{an21}
J.~An, C.~Xu, L.~Gan, and L.~Hanzo, ``Low-complexity channel estimation and
  passive beamforming for {RIS}-assisted {MIMO} systems relying on discrete
  phase shifts,'' \emph{IEEE Transactions on Communications}, pp. 1--1, 2021.

\bibitem{wei21}
L.~Wei, C.~Huang, G.~C. Alexandropoulos, C.~Yuen, Z.~Zhang, and M.~Debbah,
  ``Channel estimation for {RIS}-empowered multi-user {MISO} wireless
  communications,'' \emph{IEEE Transactions on Communications}, vol.~69, no.~6,
  pp. 4144--4157, 2021.

\bibitem{wan21}
Z.~Wan, Z.~Gao, F.~Gao, M.~D. Renzo, and M.-S. Alouini, ``Terahertz massive
  {MIMO} with holographic reconfigurable intelligent surfaces,'' \emph{IEEE
  Transactions on Communications}, vol.~69, no.~7, pp. 4732--4750, 2021.

\bibitem{an22}
J.~An, C.~Xu, Q.~Wu, D.~W.~K. Ng, M.~D. Renzo, C.~Yuen, and L.~Hanzo,
  ``Codebook-based solutions for reconfigurable intelligent surfaces and their
  open challenges,'' \emph{IEEE Wireless Communications}, pp. 1--8, 2022.

\bibitem{xu19}
J.~Xu, W.~Xu, and A.~L. Swindlehurst, ``Discrete phase shift design for
  practical large intelligent surface communication,'' in \emph{2019 IEEE
  Pacific Rim Conference on Communications, Computers and Signal Processing
  (PACRIM)}, 2019, pp. 1--5.

\bibitem{you20}
C.~You, B.~Zheng, and R.~Zhang, ``Intelligent reflecting surface with discrete
  phase shifts: Channel estimation and passive beamforming,'' in \emph{ICC 2020
  - 2020 IEEE International Conference on Communications (ICC)}, 2020, pp.
  1--6.

\bibitem{abe20}
S.~Abeywickrama, R.~Zhang, Q.~Wu, and C.~Yuen, ``Intelligent reflecting
  surface: Practical phase shift model and beamforming optimization,''
  \emph{IEEE Transactions on Communications}, vol.~68, no.~9, pp. 5849--5863,
  2020.

\bibitem{qi20}
X.~Qian, M.~Di~Renzo, J.~Liu, A.~Kammoun, and M.-S. Alouini, ``Beamforming
  through reconfigurable intelligent surfaces in single-user {MIMO} systems:
  {SNR} distribution and scaling laws in the presence of channel fading and
  phase noise,'' \emph{IEEE Wireless Communications Letters}, vol.~10, no.~1,
  pp. 77--81, 2021.

\bibitem{guo19}
H.~Guo, Y.-C. Liang, J.~Chen, and E.~G. Larsson, ``Weighted sum-rate
  maximization for intelligent reflecting surface enhanced wireless networks,''
  in \emph{2019 IEEE Global Communications Conference (GLOBECOM)}, 2019, pp.
  1--6.

\bibitem{di20}
B.~Di, H.~Zhang, L.~Song, Y.~Li, Z.~Han, and H.~V. Poor, ``Hybrid beamforming
  for reconfigurable intelligent surface based multi-user communications:
  Achievable rates with limited discrete phase shifts,'' \emph{IEEE Journal on
  Selected Areas in Communications}, vol.~38, no.~8, pp. 1809--1822, 2020.

\bibitem{mu20}
X.~Mu, Y.~Liu, L.~Guo, J.~Lin, and N.~Al-Dhahir, ``Exploiting intelligent
  reflecting surfaces in {NOMA} networks: Joint beamforming optimization,''
  \emph{IEEE Transactions on Wireless Communications}, vol.~19, no.~10, pp.
  6884--6898, 2020.

\bibitem{jun21}
M.~Jung, W.~Saad, M.~Debbah, and C.~S. Hong, ``On the optimality of
  reconfigurable intelligent surfaces ({RISs}): Passive beamforming,
  modulation, and resource allocation,'' \emph{IEEE Transactions on Wireless
  Communications}, vol.~20, no.~7, pp. 4347--4363, 2021.

\bibitem{zha21}
M.-M. Zhao, Q.~Wu, M.-J. Zhao, and R.~Zhang, ``Intelligent reflecting surface
  enhanced wireless networks: Two-timescale beamforming optimization,''
  \emph{IEEE Transactions on Wireless Communications}, vol.~20, no.~1, pp.
  2--17, 2021.

\bibitem{zha21b}
------, ``Exploiting amplitude control in intelligent reflecting surface aided
  wireless communication with imperfect {CSI},'' \emph{IEEE Transactions on
  Communications}, vol.~69, no.~6, pp. 4216--4231, 2021.

\bibitem{wu19b}
Q.~Wu and R.~Zhang, ``Beamforming optimization for wireless network aided by
  intelligent reflecting surface with discrete phase shifts,'' \emph{IEEE
  Transactions on Communications}, vol.~68, no.~3, pp. 1838--1851, 2020.

\bibitem{fu21}
M.~Fu, Y.~Zhou, Y.~Shi, and K.~B. Letaief, ``Reconfigurable intelligent surface
  empowered downlink non-orthogonal multiple access,'' \emph{IEEE Transactions
  on Communications}, vol.~69, no.~6, pp. 3802--3817, 2021.

\bibitem{hua18}
C.~Huang, G.~C. Alexandropoulos, A.~Zappone, M.~Debbah, and C.~Yuen, ``Energy
  efficient multi-user {MISO} communication using low resolution large
  intelligent surfaces,'' in \emph{2018 IEEE Globecom Workshops (GC Wkshps)},
  2018, pp. 1--6.

\bibitem{hua19}
C.~Huang, A.~Zappone, G.~C. Alexandropoulos, M.~Debbah, and C.~Yuen,
  ``Reconfigurable intelligent surfaces for energy efficiency in wireless
  communication,'' \emph{IEEE Transactions on Wireless Communications},
  vol.~18, no.~8, pp. 4157--4170, 2019.

\bibitem{han20}
Y.~Han, W.~Tang, S.~Jin, C.-K. Wen, and X.~Ma, ``Large intelligent
  surface-assisted wireless communication exploiting statistical {CSI},''
  \emph{IEEE Transactions on Vehicular Technology}, vol.~68, no.~8, pp.
  8238--8242, 2019.

\bibitem{ma20a}
X.~Ma, Z.~Chen, Y.~Chi, W.~Chen, L.~Du, and Z.~Li, ``Channel estimation for
  intelligent reflecting surface enabled terahertz {MIMO} systems,'' in
  \emph{2020 IEEE International Conference on Communications Workshops (ICC
  Workshops)}, 2020, pp. 1--6.

\bibitem{ma20b}
X.~Ma, Z.~Chen, W.~Chen, Y.~Chi, Z.~Li, C.~Han, and Q.~Wen, ``Intelligent
  reflecting surface enhanced indoor terahertz communication systems,''
  \emph{Nano Communication Networks}, vol.~24, p. 100284, 2020.

\bibitem{che19}
W.~Chen, X.~Ma, Z.~Li, and N.~Kuang, ``Sum-rate maximization for intelligent
  reflecting surface based terahertz communication systems,'' in \emph{2019
  IEEE/CIC International Conference on Communications Workshops in China (ICCC
  Workshops)}, 2019, pp. 153--157.

\bibitem{lu20}
Y.~Lu and L.~Dai, ``Reconfigurable intelligent surface based hybrid precoding
  for {THz} communications,'' \emph{arXiv preprint arXiv:2012.06261}, 2020.

\bibitem{cai20}
W.~Cai, H.~Li, M.~Li, and Q.~Liu, ``Practical modeling and beamforming for
  intelligent reflecting surface aided wideband systems,'' \emph{IEEE
  Communications Letters}, vol.~24, no.~7, pp. 1568--1571, 2020.

\bibitem{wu20}
Q.~Wu and R.~Zhang, ``Joint active and passive beamforming optimization for
  intelligent reflecting surface assisted {SWIPT} under {QoS} constraints,''
  \emph{IEEE Journal on Selected Areas in Communications}, vol.~38, no.~8, pp.
  1735--1748, 2020.

\bibitem{zha20}
Y.~Zhao, B.~Clerckx, and Z.~Feng, ``{IRS}-aided {SWIPT}: Joint waveform, active
  and passive beamforming design under nonlinear harvester model,'' \emph{IEEE
  Transactions on Communications}, pp. 1--1, 2021.

\bibitem{gon21}
S.~Gong, Z.~Yang, C.~Xing, J.~An, and L.~Hanzo, ``Beamforming optimization for
  intelligent reflecting surface-aided {SWIPT} {IoT} networks relying on
  discrete phase shifts,'' \emph{IEEE Internet of Things Journal}, vol.~8,
  no.~10, pp. 8585--8602, 2021.

\bibitem{liu22}
H.~Liu, J.~An, W.~Xu, X.~Jia, L.~Gan, and C.~Yuen, ``K-means based
  constellation optimization for index modulated reconfigurable intelligent
  surfaces,'' \emph{arXiv preprint arXiv:2207.09766}, 2022.

\bibitem{dai20}
L.~Dai, B.~Wang, M.~Wang, X.~Yang, J.~Tan, S.~Bi, S.~Xu, F.~Yang, Z.~Chen,
  M.~D. Renzo, C.-B. Chae, and L.~Hanzo, ``Reconfigurable intelligent
  surface-based wireless communications: Antenna design, prototyping, and
  experimental results,'' \emph{IEEE Access}, vol.~8, pp. 45\,913--45\,923,
  2020.

\bibitem{dun20}
M.~Dunna, C.~Zhang, D.~Sievenpiper, and D.~Bharadia, ``{ScatterMIMO}: Enabling
  virtual {MIMO} with smart surfaces,'' in \emph{Proceedings of the 26th Annual
  International Conference on Mobile Computing and Networking}, 2020, pp.
  1--14.

\bibitem{she20}
S.~Shen, B.~Clerckx, and R.~Murch, ``Modeling and architecture design of
  reconfigurable intelligent surfaces using scattering parameter network
  analysis,'' \emph{IEEE Transactions on Wireless Communications}, pp. 1--1,
  2021.

\bibitem{xu21b}
J.~Xu, Y.~Liu, X.~Mu, and O.~A. Dobre, ``{STAR-RISs}: Simultaneous transmitting
  and reflecting reconfigurable intelligent surfaces,'' \emph{IEEE
  Communications Letters}, vol.~25, no.~9, pp. 3134--3138, 2021.

\bibitem{li22-1}
H.~Li, S.~Shen, and B.~Clerckx, ``Beyond diagonal reconfigurable intelligent
  surfaces: From transmitting and reflecting modes to single-, group-, and
  fully-connected architectures,'' \emph{IEEE Transactions on Wireless
  Communications}, pp. 1--1, 2022.

\bibitem{li22-2}
------, ``Beyond diagonal reconfigurable intelligent surfaces: A multi-sector
  mode enabling highly directional full-space wireless coverage,'' \emph{IEEE
  J. Sel. Areas Commun.}, pp. 1--1, 2023.

\bibitem{li22-3}
------, ``A dynamic grouping strategy for beyond diagonal reconfigurable
  intelligent surfaces with hybrid transmitting and reflecting mode,''
  \emph{IEEE Trans. Veh. Technol.}, pp. 1--6, 2023.

\bibitem{li22}
Q.~Li, M.~El-Hajjar, I.~Hemadeh, A.~Shojaeifard, A.~A.~M. Mourad, B.~Clerckx,
  and L.~Hanzo, ``Reconfigurable intelligent surfaces relying on non-diagonal
  phase shift matrices,'' \emph{IEEE Transactions on Vehicular Technology},
  vol.~71, no.~6, pp. 6367--6383, 2022.

\bibitem{poz11}
D.~M. Pozar, \emph{Microwave engineering}.\hskip 1em plus 0.5em minus
  0.4em\relax John wiley \& sons, 2011.

\bibitem{cle13}
B.~Clerckx and C.~Oestges, \emph{{MIMO} wireless networks: Channels, techniques
  and standards for multi-antenna, multi-user and multi-cell systems}.\hskip
  1em plus 0.5em minus 0.4em\relax Academic Press, 2013.

\bibitem{wu19c}
Q.~Wu and R.~Zhang, ``Intelligent reflecting surface enhanced wireless network
  via joint active and passive beamforming,'' \emph{IEEE Transactions on
  Wireless Communications}, vol.~18, no.~11, pp. 5394--5409, 2019.

\bibitem{zap21}
A.~Zappone, M.~Di~Renzo, F.~Shams, X.~Qian, and M.~Debbah, ``Overhead-aware
  design of reconfigurable intelligent surfaces in smart radio environments,''
  \emph{IEEE Transactions on Wireless Communications}, vol.~20, no.~1, pp.
  126--141, 2021.

\bibitem{aud02}
C.~Audet and J.~E. Dennis~Jr, ``Analysis of generalized pattern searches,''
  \emph{SIAM Journal on optimization}, vol.~13, no.~3, pp. 889--903, 2002.

\bibitem{bis06}
C.~M. Bishop, \emph{Pattern Recognition and Machine Learning}.\hskip 1em plus
  0.5em minus 0.4em\relax Springer Verlag, Aug. 2006.

\bibitem{jae14}
S.~Jaeckel, L.~Raschkowski, K.~B{\"o}rner, and L.~Thiele, ``{QuaDRiGa}: A 3-{D}
  multi-cell channel model with time evolution for enabling virtual field
  trials,'' \emph{IEEE Transactions on Antennas and Propagation}, vol.~62,
  no.~6, pp. 3242--3256, 2014.

\end{thebibliography}
